\documentclass[aps,prd,twocolumn,showpacs,eqsecnum,amssymb,amsmath,showpacs,a4paper,superscriptaddress,nofootinbib]{revtex4-1}
\usepackage{graphicx,multirow}
\usepackage{color}
\usepackage{amsfonts}
\usepackage{mathrsfs}
\usepackage[latin1]{inputenc}      		% tradução de acentos
\usepackage{booktabs}
\usepackage{longtable}

\def\be{\begin{equation}}
\def\ee{\end{equation}}
\def\ba{\begin{align}}
\def\ea{\end{align}}

\begin{document}
%--------------------------------------------------------------
\setcounter{topnumber}{1}

\title{Quasinormal modes of charged fields around a Reissner-Nordstr\"om black hole}
%--------------------------------------------------------------
\author{Maur\'icio Richartz}
\email{mauricio.richartz@ufabc.edu.br}
\author{Davi Giugno}
\email{dgiugno@if.usp.br}
\affiliation{Centro de Matem\'atica, Computa\c{c}\~ao e Cogni\c{c}\~ao, Universidade Federal do ABC (UFABC), 09210-170 Santo Andr\'e, SP, Brazil}
%--------------------------------------------------------------
%------------------------------------------------------------------------------------------------------------------------------------------
%------------------------------------------------------------------------------------------------------------------------------------------
%------------------------------------------------------------------------------------------------------------------------------------------
%------------------------------------------------------------------------------------------------------------------------------------------
%
%------------------------------------------------------------------------------------------------------------------------------------------

%--------------------------------------------------------------
\begin{abstract}
%--------------------------------------------------------------

The quasinormal spectrum of a charged scalar field around a non-extremal Reissner-Nordstr\"om black hole, specially in the limit of large electromagnetic interaction, has been comprehensively studied only very recently. In this work, we extend the analysis to Dirac fields using the continued fraction method and compare the results with the scalar case. In particular, we study the behaviour of the fundamental quasinormal mode as a function of the black hole's charge and of the electromagnetic interaction parameter. We derive an analytical formula for the quasinormal frequencies in the limit of large electromagnetic interaction. As the extremal limit of black hole charge is approached, we show that, unlike the case of neutral fields, the imaginary part of the quasinormal frequencies approach zero for charged fields.

%--------------------------------------------------------------
\end{abstract}
%--------------------------------------------------------------
\maketitle

%================
\section{Introduction}
%================

The study of black hole perturbations was pioneered by Regge and Wheeler in the 1950s while attempting to establish the stability of Schwarzschild black holes~\cite{regge}. Later, Vishveshwara~\cite{vish} identified a special type of perturbation characterized by purely outgoing waves at spatial infinity and purely ingoing waves in the vicinity of the event horizon. Such perturbations, dubbed quasinormal modes (QNMs) by Press~\cite{press}, are only permitted for a discrete set of complex frequencies $\omega _ n$, the so--called quasinormal frequencies. Assuming a time dependence of $\exp(-i\omega t)$, if their imaginary part is negative ($Im(\omega_n) < 0 $), the perturbation is damped; similarly, if $Im(\omega_n) > 0 $, the mode is dynamically unstable and its amplitude grows in time until the linear approximation ceases to be valid.

QNMs have been studied in many different contexts, including (but not limited to) the scattering of particles by black holes~\cite{det1,det2,davis}, the formation of black holes by the collapse of a star~\cite{cunn1,cunn2}, the stability of black holes~\cite{teuko,whiting}, the quantization of black holes~\cite{hodquantum, hodquantum2, corda}, the AdS/CFT correspondence~\cite{ads}, and the perturbation of analogue black holes~\cite{ana}. Perhaps the most famous manifestation of QNMs occurs in Astrophysics, when gravitational waves are emitted by a perturbed black hole. Such process can be divided into three stages. First, there is an outburst of radiation which is highly dependent on the initial perturbation; second, there is a long period of quasinormal oscillation, and finally, at later times, the QNMs damp out and become dominated by a power law tail. 
 
According to the uniqueness theorems, once a black hole has reached a stable configuration, it will be characterized only by three parameters: its mass $M$, its charge $Q$, and its angular momentum $J$. The QNM $\omega_n$ of a black hole, on the other hand, will in general be described by an interacting field which is characterized by a mass parameter $\mu$, a charge parameter $q$, an orbital number $\ell$, an azimuthal number $m$ and an spin parameter $s$ (depending on the type of perturbation, i.e.~scalar, Dirac, electromagnetic or gravitational). There exists in the literature an extensive amount of research on QNMs, for almost any kind of black hole and any kind of perturbation -- see e.g.~\cite{review1,review2,review3,review4} for recent reviews on the subject. There is one situation, however, which only recently has drawn the attention of the physics community. The QNMs of charged fields in a charged black hole background have only been studied in detail in the regime $\mu M \ll 1 $ and $qQ \ll 1$~\cite{konoa,konob,zhid,chang,jing,wu}, the exception being Refs.~\cite{hodap,hod,konoplya}, where the scalar case was analysed. In view of that, our main objective in this work is to use the continued fraction method to present a detailed study of the quasinormal modes of a massless charged Dirac field for arbitrary values of $qQ$, comparing the obtained results with the scalar case. We also discuss in detail the behaviour of the QNM frequencies in the limit $Q\rightarrow M$ of extremal black holes.

%================
\section{Field dynamics}
%================

Let us start by describing the dynamics of massless charged fields around a Reissner-Nordstr\"om (RN) black hole of mass $M$ and charge $Q$ whose metric is given by
\be
\label{RN}
ds^2 = - \frac{\Delta}{r^2} dt^2 + \frac{r^2}{\Delta} dr^2 + r^2 \left( d \theta ^2 + \sin ^2 \theta  d \varphi ^2 \right),
\ee
where $\Delta = r^2 - 2Mr + Q^2$. The locations of the event horizon and of the Cauchy horizon are, respectively, $r_+=M+\sqrt{M^2-Q^2}$ and $r_-=M-\sqrt{M^2-Q^2}$. The RN metric can be obtained from the Einstein-Maxwell equations by assuming stationarity and spherical symmetry. The electromagnetic field is obtained from the electromagnetic potential $A_ \mu$, whose only non-vanishing component is $A_0= -Q/r$.

A massless charged scalar field $\psi$ propagating on a RN background obeys the Klein-Gordon equation
\be \label{kg0}
%--
 \left( \nabla^\mu - ie A^{\mu} \right)\left( \nabla_\mu - ie A_{\mu} \right)  \, \psi = 0.
%--
\ee
A massless charged Dirac field, on the other hand, can be described by a pair of spinors $P^A$ and $\bar{Q}^A$ which satisfy the Dirac equations,
\begin{align}
\sigma^ \mu {}_{AB'} \left( \nabla _\mu - i q A_\mu \right) P^A  &= 0, \label{dirrn1} \\
\sigma^ \mu {}_{AB'} \left( \nabla _\mu + i q A_\mu \right) Q^A  &= 0, \label{dirrn2}
\end{align} 
where $\sigma^\mu {}_{AB'}$ are generalizations of the Pauli matrices~\cite{stewart}. Note that, by using the differential operator $\nabla _\mu - i q A_\mu$, we have assumed the minimal coupling between the scalar/Dirac field and the electromagnetic field of the black hole. 

All the equations above can be separated if one uses appropriate ansatzes. In fact, let us rewrite the scalar field as $\psi = R_0(r) Y_{j m}^0 (\theta, \phi) e^{- i \omega t}$ and the fermionic spinors as 
\begin{align}
P^0 &= \frac{ R_{-\frac{1}{2}}(r)Y_{j m}^{-\frac{1}{2}}(\theta)}{ r } e^{- i \omega t} e ^{i m \varphi} , \\
P^1 &= R_{+\frac{1}{2}}(r)Y_{j m}^{+\frac{1}{2}}(\theta) e^{- i \omega t} e ^{i m \varphi} , \\ 
\bar Q ^{1'} &= R_{+\frac{1}{2}}(r) Y_{j m}^{-\frac{1}{2}}(\theta)e^{- i \omega t} e ^{i m \varphi} , \\
\bar Q ^{0'} &= - \frac{ R_{-\frac{1}{2}}(r)Y_{j m}^{+\frac{1}{2}}(\theta)}{r} e^{- i \omega t} e ^{i m \varphi},
\end{align}
where $\omega$ is the frequency of the field, $j \ge |s|$ and $-j \le m \le j$ are integers (scalar case) or half-integers (fermionic case) and $Y_{j m}^{s}(\theta)$ are the corresponding spin-weighted spherical harmonics~\cite{goldberg} (note that for scalar fields, $\ell$ is usually used instead of $j$). After some algebra, one can show that the radial components of the fields, $R_{s}(r)$, satisfy the following master equation~\cite{jing,mauricio_alberto}, 
\begin{align}
  \Delta^{-s} \frac{d}{dr} \left(\Delta^{s+1}  \frac{dR_{s} }{dr}  \right)  +
\left[
\frac{K^2 -  2is (r-M)K}{\Delta} \right. \nonumber \\ 
 + 4is \omega r - 2isqQ - \lambda _s \bigg] R_s   =0, \label{master}
\end{align}
where $K=\omega r^2 -qQr$ and $\lambda _s = (j - s)(j + s + 1)$ is a separation constant. The equation above is analogous to the radial Teukolsky equation for the Kerr metric~\cite{teukolsky}.

If we now make the transformation $f_s=\Delta ^{s/2}rR_{s}$ and adopt the tortoise coordinate $r_*$ (defined by $dr_* / dr = r^2 / \Delta$), the master equation~\eqref{master} becomes
\be \label{eq3}
\frac{d^2f_s}{dr_* ^2} +  W_s(\omega,r_*) f_s=0,
\ee
where the complex function $W_s$ is given by
\begin{align} \label{pot}
W_s(\omega,r_*)=\frac{\Delta}{r^4}\left[ \frac{\left( K - i s (r - M) \right) ^2}{\Delta}  + 4 is \omega r \right. \nonumber \\
\left.   - 2 isqQ - j(j+1) + s^2 - 2 \frac{M}{r} + 2 \frac{Q^2}{r^2}  \right].
\end{align}

The equation above can be solved analytically in the asymptotic limits of the tortoise coordinate space (note that $r_* \rightarrow -\infty$ and $r_* \rightarrow \infty$ correspond, respectively, to the event horizon $r \rightarrow r_+$ and the spatial infinity $r\rightarrow \infty$). By further imposing the boundary conditions naturally associated with quasinormal oscillation, i.e.~purely outgoing waves far away from the black hole and purely ingoing waves near the black hole's event horizon, one is able to obtain the asymptotic form of the quasinormal modes. For a non-extremal black hole ($0\le Q < M$), the result is
\be \label{asymp}
f_s \rightarrow \begin{cases} Z_s^{\rm out} r_* ^{-s - i q Q} e^ {+ i \omega r_*},   &\quad r_* \rightarrow \infty \\
                 Z_s^{\rm tr} e^{-\frac{s}{2}\left(\frac{r_+ - r_-}{r_+ ^2} \right)r_* - i \left( \omega - \frac{q Q}{r_+} \right) r_ * }, &\quad r_* \rightarrow -\infty
                 \end{cases}
\ee
where $Z_s^{\rm out}$ and $Z_s^{\rm tr}$ are constants.

Equation~\eqref{eq3}, together with the boundary conditions above, becomes an eigenvalue problem for $\omega$ so that only a discrete set of frequencies (the quasinormal frequencies $\omega_n$) is allowed. In the next section we describe the continued fraction method, a commonly used method to determine the quasinormal frequencies, and discuss its implementation for charged fields around a RN black hole. 

%================
\section{Continued Fractions}
%================

In 1985, inspired by a technique due to Jaffe to calculate the energy eigenvalues of the $H_2^{+}$ ion, Leaver devised a numerical method to compute QNMs of Schwarzschild and Kerr black holes using continued fractions~\cite{leaver1,leaver2}. His approach consists in expressing the solution to the wave equation, with the appropriate boundary conditions for QNMs, as a power series which is everywhere convergent, except possibly at the asymptotic limit $r \rightarrow \infty$. By requiring convergence also at infinity, one obtains a continued fraction relation which must be satisfied by the expansion coefficients of the power series.

Leaver's original method was later improved by Nollert~\cite{nollert} and has since been used to calculate QNMs in a variety of situations~\cite{review4}. Recently, it has been used in Ref.~\cite{konoplya} to obtain the QNMs of a charged scalar field around a Kerr--Newman black hole. In this paper, besides the scalar case, we also apply  Leaver's technique to determine the QNMs of a massless Dirac field around a RN black hole without any restrictions on the parameters involved. To the best of our knowledge, it is the first time that these QNMs are calculated for Dirac fields without the assumption of small $qQ$. For small $qQ$, the QNMs of charged Dirac fields around a RN black hole have been calculated in Refs.~\cite{chang} and~\cite{wu} using, respectively, the WKB and the P\"oschl--Teller methods. We also note that the continued fraction method has been applied to the more general context of a Kerr-Newman-de Sitter background in Ref.~\cite{zhid}.

To implement Leaver's method, we start by noting that, in the non-extremal case, eq.~\eqref{master} has an irregular singularity at $r=+\infty$ and three regular singularities at $r=0$, $r=r_-$ and $r=r_+$. The solution which satisfies the boundary conditions~\eqref{asymp} can be expanded in a powers series around $r=r_+$,
\begin{align} \label{exp}
R(r)=e^{i\omega r}(r-r_-)^{\epsilon}\sum_{n=0}^{\infty}a_n\left(\frac{r-r_+}{r-r_-} \right)^{n + \delta},
\end{align}
where $\epsilon=-i qQ + i \omega  (r_+ + r_-) -2 s-1$ and $\delta=-s-\frac{i r_+^2 \left(\omega -\frac{qQ}{r_+}\right)}{r_+-r_-}$.
The coefficients $a_n$ must satisfy the three-term recurrence relations
\be \label{rec0}
\begin{cases} 
\alpha_0 a_1 + \beta_0 a_0 = 0, \\
\alpha_n a_{n+1} + \beta_n a_n + \gamma_n a_{n-1} = 0, \qquad n \ge 1 \\
\end{cases}
\ee 
where $\alpha_n$, $\beta_n$ and $\gamma_n$ are given by 
\begin{widetext}
\begin{align}
\alpha_n &=  -(n+1) \left[ r_- (n-s+1)+ r_+ (-n-2 i qQ +2 i r_+ \omega +s-1) \right] , \label{al} \\
\beta_n  &= - r_+ \left[\lambda_s + 2 n^2-4 i r_+ \omega  (2 n+3 i qQ +1)+6 i n qQ +2 n-4 (qQ)^2+3 i qQ-8 r_+^2 \omega ^2+s+1\right]  \nonumber \\
& \qquad \qquad   +r_  - \left[\lambda_s +2 n (n+i qQ +1)+i qQ +s+1\right]-2 i (2 n+1) r_- r_+ \omega , \label{be}\\
\gamma_n &= -\left\{n+2 i \left[qQ-\omega  (r_+ + r_-)\right]\right\}\left[n (r_- - r_+)+i r_+ (-2 qQ +2 r_+ \omega +i s)+ r_- s\right] \label{ga}  .
\end{align}
\end{widetext}
When $q=Q=s=0$, these coefficients reduce to the original ones obtained by Leaver for the Schwarzschild case~\cite{leaver1}. Additionally, for uncharged fields ($qQ=0$), these coefficients are compatible with the ones obtained in Ref.~\cite{jing2}; for scalar fields ($s=0$), they are compatible with the ones obtained in Ref.~\cite{hodquantum2}.

Since $r_+$ is a regular singular point, convergence of the series is automatically guaranteed for $r_+ \le r < \infty$. Convergence also at $r=\infty$ implies convergence of the sum $\sum_n a_n$ and, therefore, the coefficients $a_n$ must satisfy the following infinite continued fraction equation~\cite{gaut},
\be \label{equat1}
0 = \beta_0 - \frac{\alpha_0\gamma_1}{\beta_1 -}\frac{\alpha_1\gamma_2}{\beta_2 -} \dots
\ee
Any QNM can be found by solving the equation above if the infinite continued fraction is truncated at some sufficiently large index N. However, in practice, the equation above is only used to calculate the fundamental frequency (its most stable root). In order to find the $n$-th quasinormal frequency, it is more convenient to invert the equation above $n$-times, resulting in the following equation,
\be \label{cfeqn}
\beta_n -  \frac{\alpha_{n-1}\gamma_n}{\beta_{n-1} -}\frac{\alpha_{n-2}\gamma_{n-1}}{\beta_{n-2} -}\dots \frac{\alpha_0\gamma_1}{\beta_0} =  \frac{\alpha_{n}\gamma_{n+1}}{\beta_{n+1} -}\frac{\alpha_{n+1}\gamma_{n+2}}{\beta_{n+2} -}\dots,
\ee
whose most stable root is exactly the $n$-th QNM frequency.  

As discovered by Nollert~\cite{nollert} for Schwarzschild black holes, the convergence of Leaver's method can be improved if one estimates the `error' associated with the truncation of the continued fraction. To implement the technique in our problem, we first note that $R_N=-a_{N+1}/a_N$ satisfies the recursive equation
\be \label{rec22}
R_N=\frac{\gamma_{N+1}}{\beta_{N+1} - \alpha_{N+1}R_{N+1}},
\ee  
and, therefore, corresponds exactly to the `rest' of the continued fraction~\eqref{equat1} truncated at order N. We then expand $R_N$ in a power series of $N^{1/2}$,
\be \label{expan}
R_N= \sum_{k=0}^{\infty} C_k N^{-k/2},
\ee
and substitute it in eq.~\eqref{rec22} to determine the coefficients $C_k$. The first coefficients obtained this way are given by $C_0=-1$, $C_1=\sqrt{2i\omega(r_--r_+)}$ and $C_2=\frac{3}{4}-s-iqQ+2i\omega r_+$.
% $C3=\frac{16 \lambda_s -16 (qQ)^2+64 qQ r_+ \omega +16 i (4 s-5) \omega  (r_- -r_+)-64 r_+^2 \omega ^2+16 s^2+16 s+3}{32 C1}$

%================
\section{Numerical Results}
%================

We shall now use a root--finding algorithm to determine the QNMs associated with equations~\eqref{equat1} and~\eqref{cfeqn}. It is convenient to rescale the equations so that the parameters and variables become dimensionless. Indeed, if we rescale $r \rightarrow r/M$ and $t \rightarrow Mt$, it is straightforward to see that the $n$-th quasinormal frequency $M\omega_n$ will depend only on the following parameters: $Q/M$, $qQ$, $s$ and $j$. We focus our analysis in the case $s=j=1/2$ and investigate the behaviour of the fundamental ($n=0$) QNMs as $Q/M$ and $qQ$ are varied. As explained in Refs.~\cite{wu,chang}, the potentials for $s=\pm 1/2$ produce the same QNM spectrum and, therefore, there is no loss of generality in considering only the $s=+1/2$  case. As mentioned before, we pay special attention to the regime $qQ \gg 1$ and also discuss in detail what happens when the limit $Q \rightarrow M$ of extremal charge is approached. For comparison reasons, we have also calculated the fundamental QNMs for the scalar case $s=j=0$.  The results for different sets of parameters are presented in table~\ref{table1}.

\begin{table*}
\centering \caption{QNM frequencies of the fundamental mode obtained with the continued fraction method for different values of $Q/M$ and $qQ$. For each value of $qQ$, the first line corresponds to the scalar case $s=j=0$ and the second line to the fermionic case $s=j=1/2$.} \vskip 12pt
\begin{tabular}{@{}c|cccccccccc@{}}
\hline 
&\multicolumn{2}{c}{$Q/M=0.01$}&\multicolumn{2}{c}{$Q/M=0.1$}&\multicolumn{2}{c}{$Q/M=0.5$}&\multicolumn{2}{c}{$Q/M=0.99$}&\multicolumn{2}{c}{$Q/M=0.9999$}\\ \hline
$|qQ|$       & $|\text{Re}(\omega)|$     &$\text{Im}(\omega)$ & $|\text{Re}(\omega)|$ &$\text{Im}(\omega)$ & $|\text{Re}(\omega)|$ &$\text{Im}(\omega)$ & $|\text{Re}(\omega)|$ &$\text{Im}(\omega)$ & $|\text{Re}(\omega)|$ &$\text{Im}(\omega)$ \\
 \multirow{2}{*}{$0$}   & $0.110457$ & $- 0.104896 $  &  $ 0.110649$ & $ - 0.104938$  & $ 0.115764$ & $ - 0.105751 $ & $ 0.133570$ & $ - 0.095641 $ & $0.133459$ &  $- 0.095844$  \\
           & $ 0.182966$ & $- 0.096983 $  &  $ 0.183295$ & $ - 0.097033$  & $ 0.192120$ & $ - 0.098106 $ & $ 0.236845$ & $ - 0.088925 $ & $ 0.238169$ &  $- 0.087697$ \\
 \multirow{2}{*}{$0.01$}   & $0.106575$ & $-0.104213 $  &  $0.106760$ & $ -0.104254$  & $0.111673$ & $-0.105068 $ & $0.138611$ & $-0.095570 $ &  $0.010010$ &  $-0.014155$   \\
           & $0.179503$ & $-0.096446 $  &   $0.179824$ & $-0.096496$  & $0.188446$ & $-0.097572 $ & $0.231900$ & $ -0.088850 $ & $0.010008$ &  $-0.021229$ \\
   \multirow{2}{*}{$0.1$}   & $0.072255$ & $-0.096978 $  &  $0.072377$ & $ -0.097016$  & $0.075611$ & $-0.097778 $ & $0.186342$ & $-0.091844 $ & $0.100110$ &  $-0.014005$   \\
           & $0.149158$ & $-0.091157 $  & $ 0.149416$ & $-0.091206$  & $0.156303$ & $ -0.092289 $ & $0.189365$ & $-0.087148 $ & $0.100081$ &  $-0.021159$ \\
            \multirow{2}{*}{$1$}   & $0.547816$ & $-0.125258 $  &  $0.549058$ & $ -0.125264$  & $0.583393$ & $ - 0.124817 $ & $0.895177$ & $ - 0.054159 $ &  $0.987906$ &  $- 0.006805$   \\
           & $0.590192$ & $- 0.120632 $  &  $0.591511$ & $ - 0.120655$  & $0.627858$ & $ - 0.120744 $ & $0.933299$ & $ -0.058255 $ & $0.996092$ &  $-0.008213$ \\
            \multirow{2}{*}{$10$}   & $5.006341$ & $- 0.125074 $  &  $5.018771$ & $ - 0.125073$  & $5.364954$ & $ - 0.124410 $ & $8.765415$ & $ - 0.054161 $ &  $9.860731$ &  $- 0.006875$   \\
           & $5.012548$ & $- 0.124996 $  &  $5.024987$ & $ - 0.124996$  & $5.371354$ & $ - 0.124353 $ & $8.769147$ & $ - 0.054172 $ &  $9.861243$ &  $-0.006875$  \\ 
            \multirow{2}{*}{$100$}   & $50.00188$ & $- 0.125001 $  &  $50.12625$ & $ - 0.125000$  & $53.59044$ & $ - 0.124356 $ & $87.63741$ & $ - 0.054172 $ &  $98.60556$ &  $-0.006875$   \\
           & $50.00250$ & $- 0.125000 $  & $50.12688$ & $ - 0.124999$  & $53.59108$ & $ -0.124356 $ & $87.63779$ & $ -0.054172 $ & $98.60561$ &  $ -0.006875$ \\
            \multirow{2}{*}{$1000$}   & $500.0126$ & $- 0.125000 $  &  $501.2564$ & $ - 0.124999$  & $535.8984$ & $ - 0.124356 $ & $876.3725$ & $ - 0.054172 $ & $986.0554$ &  $- 0.006875$   \\
           & $500.0126$ & $-0.125000 $  &  $501.2564$ & $ - 0.124999$  & $535.8985$ & $ -0.124356 $ & $876.3725$ & $ - 0.054172 $ & $986.0554$ &  $- 0.006875$ \\       
\hline 
\end{tabular}
\label{table1}
\end{table*}  

\begin{figure*}
\begin{center}
\begin{tabular}{cc}
\includegraphics[width=8.6cm]{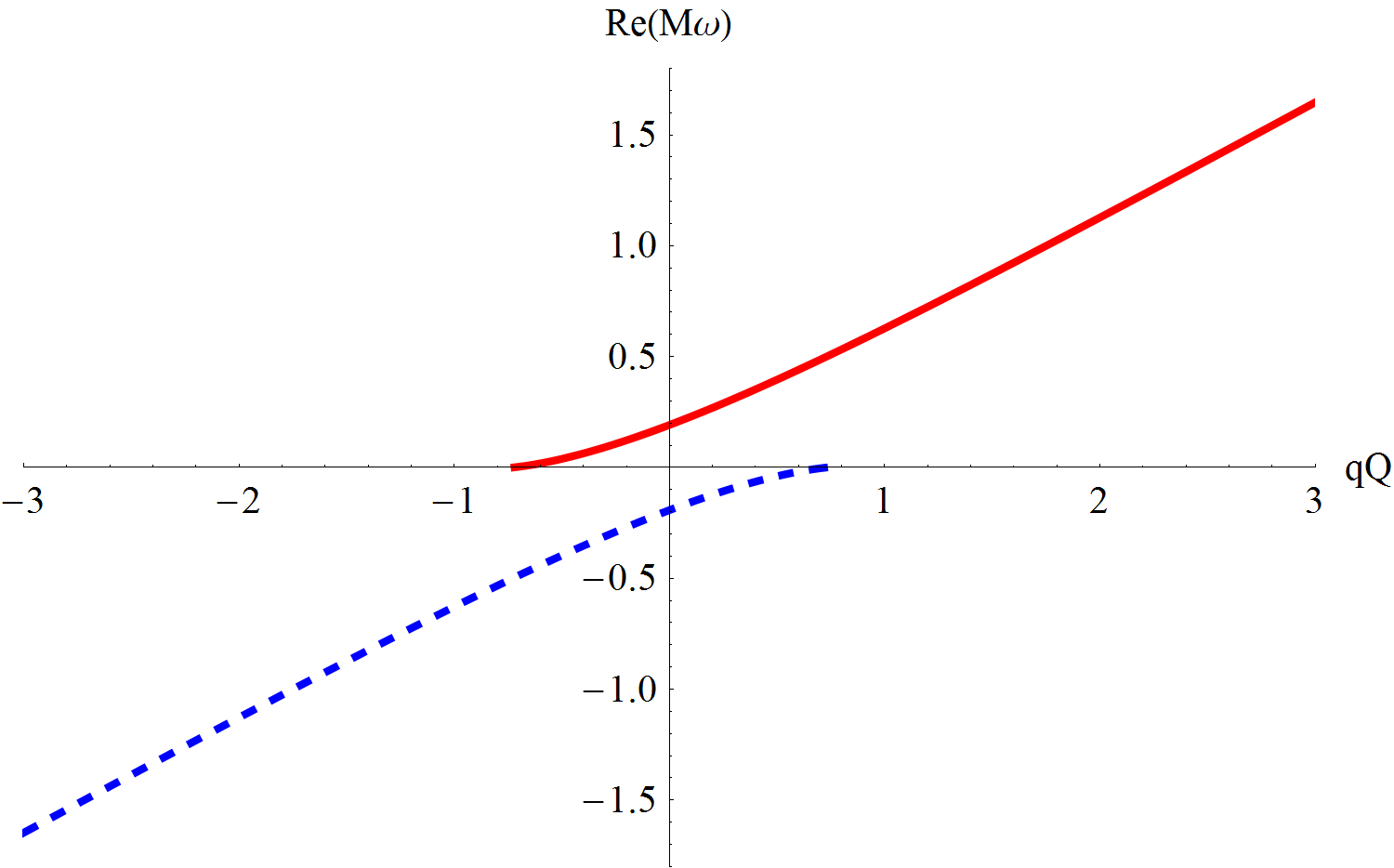}&
\includegraphics[width=8.6cm]{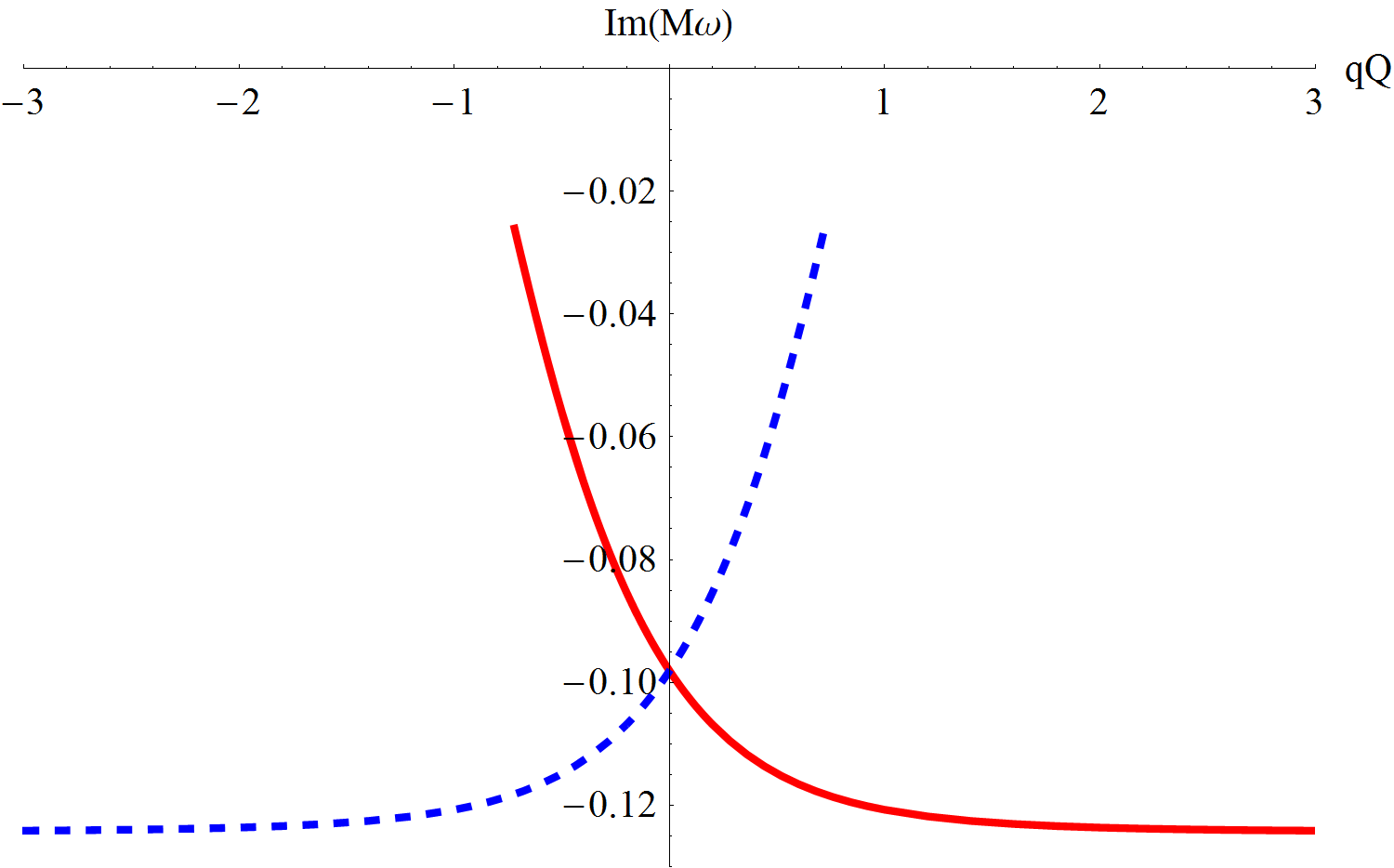}\\
\end{tabular}
\caption{Real (left plot) and imaginary (right plot) parts of the most stable QNMs of the Dirac field $s=j=1/2$ as a function of $qQ$ for $Q/M=0.5$. Note the symmetry with respect to the simultaneous transformations $\omega \rightarrow -\omega^*$ and $qQ \rightarrow - qQ$. Note also the critical value of $|qQ|$ above which one of the branches disappears from the spectrum. 
\label{figa}
}
\end{center}
\end{figure*}

In order to get a complete description of the dependence of the quasinormal modes on the electromagnetic coupling between the field and the black hole, we plot the real and imaginary parts of the fundamental frequency as a function of $qQ$. The results are presented in Fig.~\ref{figa} for the case $Q/M=0.5$. In fact, in order to highlight the symmetry of the quasinormal modes with respect to the transformation ($\omega \rightarrow -\omega^*, \quad qQ \rightarrow - qQ$), we have plotted both the $Re(\omega)>0$ and the $Re(\omega)<0$ quasinormal branches in Fig.~\ref{figa}.  For sufficiently small values of $|qQ|$, when $qQ$ is positive, the mode with $Re(\omega)<0$ is, in general, more stable than the mode with $Re(\omega)>0$ (because of the symmetry, this behaviour is reversed for negative $qQ$). As $|qQ|$ increases, the real part of the fundamental frequency approaches zero until its branch ceases to exist and disappears from the spectrum at some critical value of the electromagnetic coupling. Such behaviour, which is not uncommon, was observed in Ref.~\cite{konoplya} for scalar fields. We have now verified that it is also manifest for fermionic fields. As this critical value is approached, the continued fraction method seems to converge slower. Nonetheless, we were able to verify that this critical value is almost unchanged as $Q/M$ varies: for the scalar case ($s=j=0$) it is $|qQ|\approx 0.3$ and for the Dirac case ($s=j=1/2$) it is $|qQ| \approx 0.7$. 

\begin{figure*}
\begin{center}
\begin{tabular}{cc}
\includegraphics[width=8.6cm]{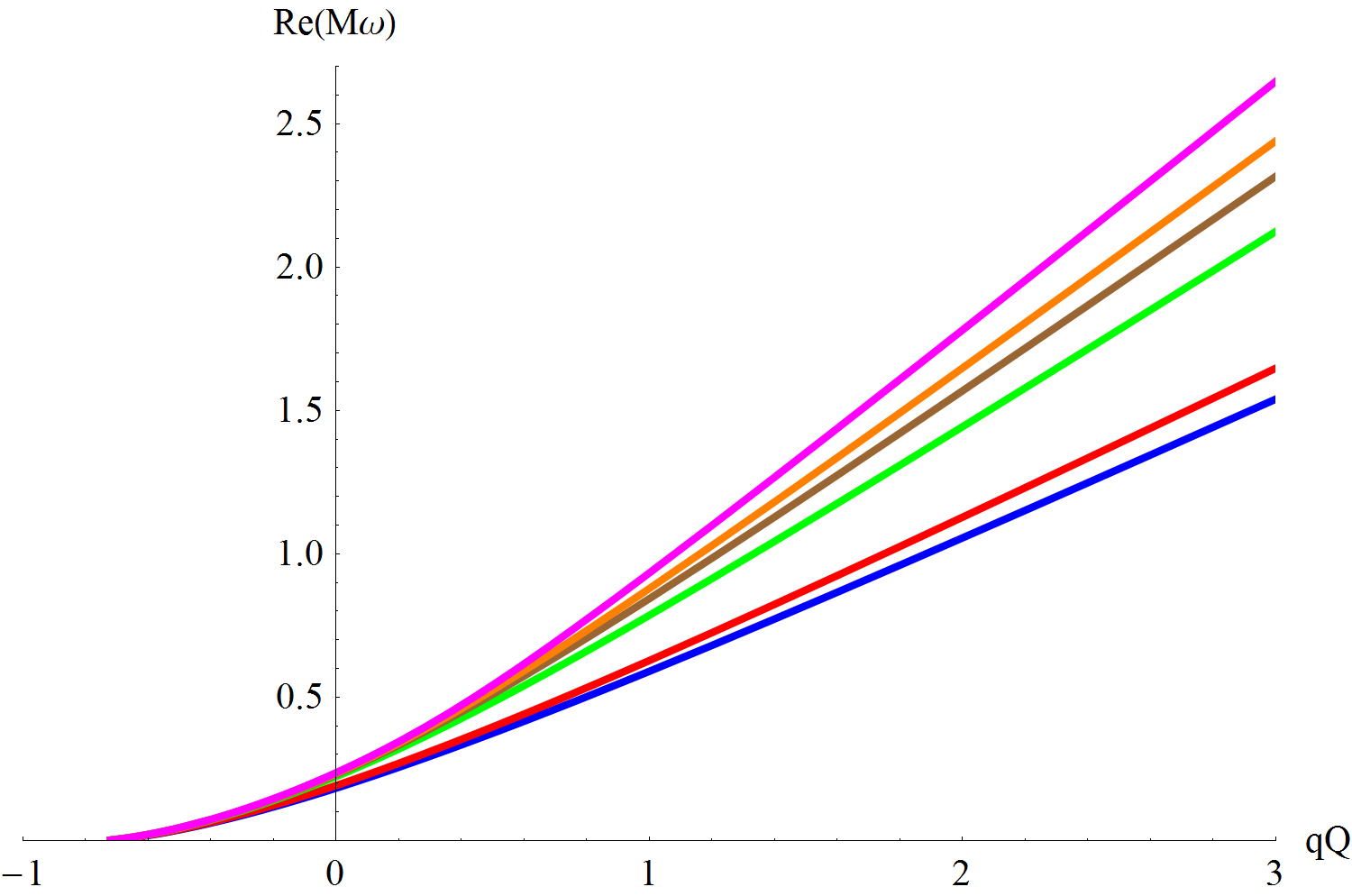}&
\includegraphics[width=8.6cm]{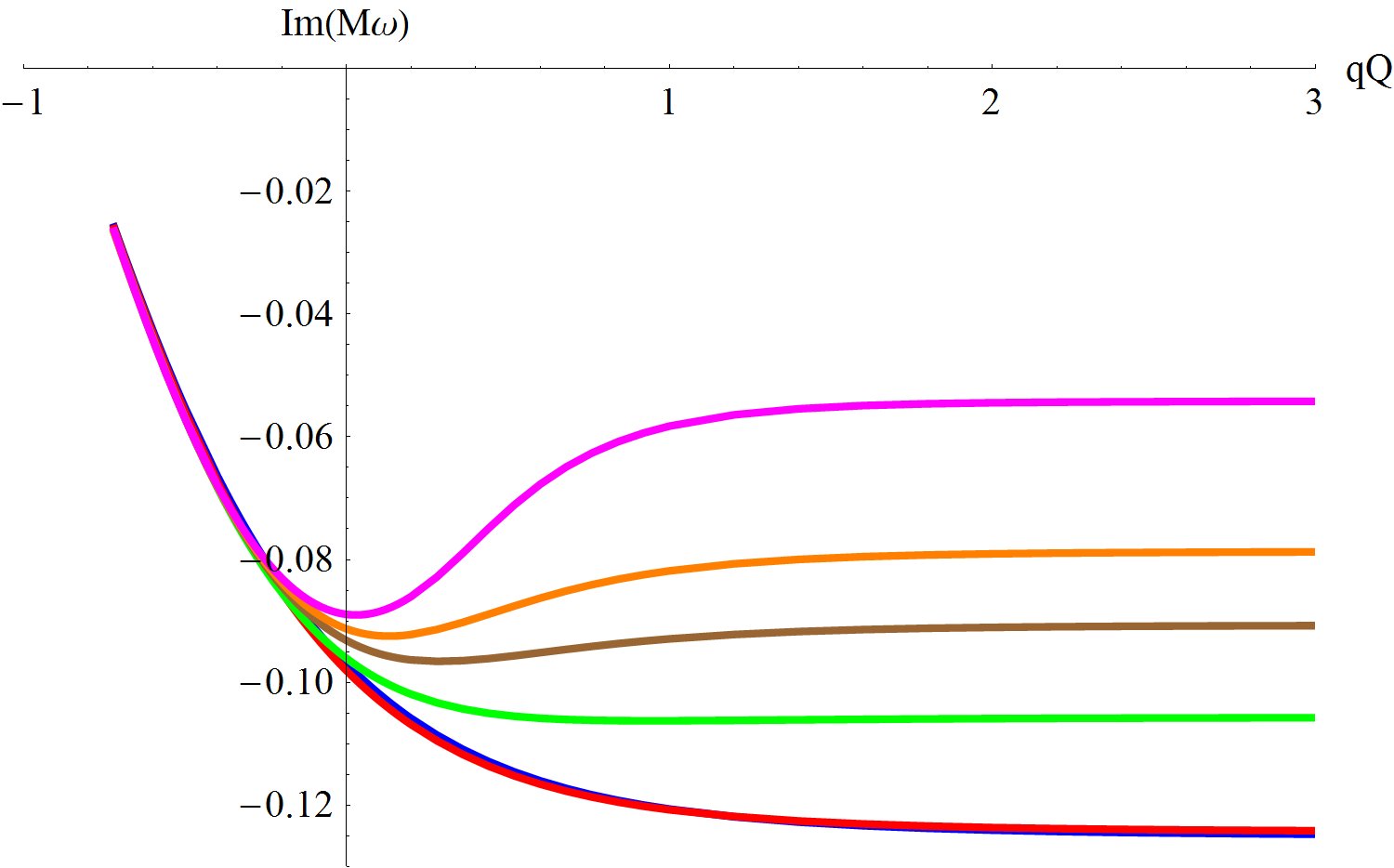}\\
\end{tabular}
\caption{Real (left plot) and imaginary (right plot) parts of the fundamental quasinormal mode as a function of $qQ$ for the fermionic case $s=j=1/2$ (only the $Re(\omega)>0$ branch is represented here). From bottom to top: $Q/M=0.01$ (blue), $Q/M=0.5$ (red), $Q/M=0.9$ (green), $Q/M=0.95$ (brown), $Q/M=0.97$ (orange), $Q/M=0.99$ (magenta). Note that the imaginary parts for $Q/M=0.01$ and $Q/M=0.5$ are almost indistinguishable in this plot. 
\label{figb}
}
\end{center}
\end{figure*}

\begin{figure*}
\begin{center}
\begin{tabular}{cc}
\includegraphics[width=8.6cm]{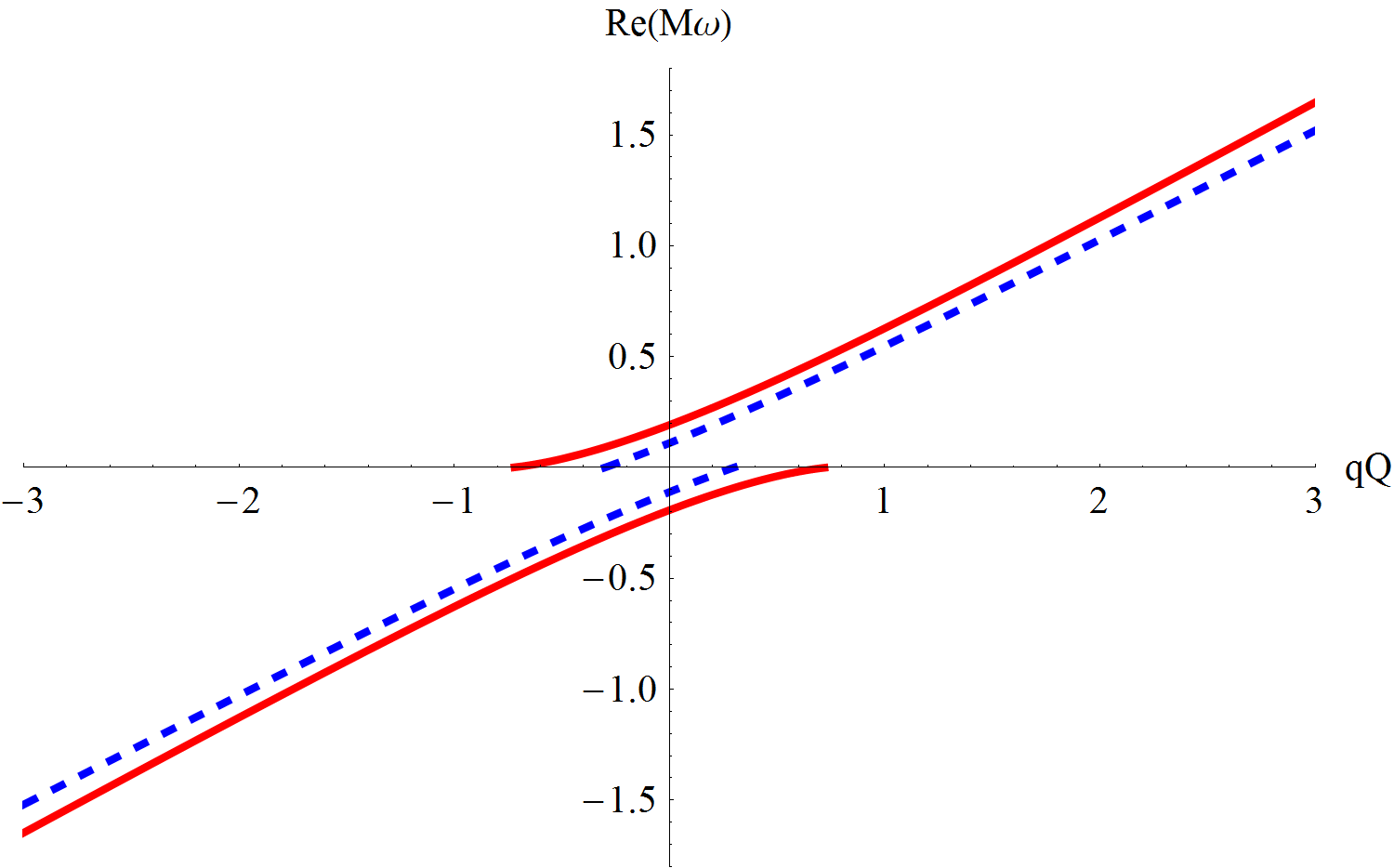}&
\includegraphics[width=8.6cm]{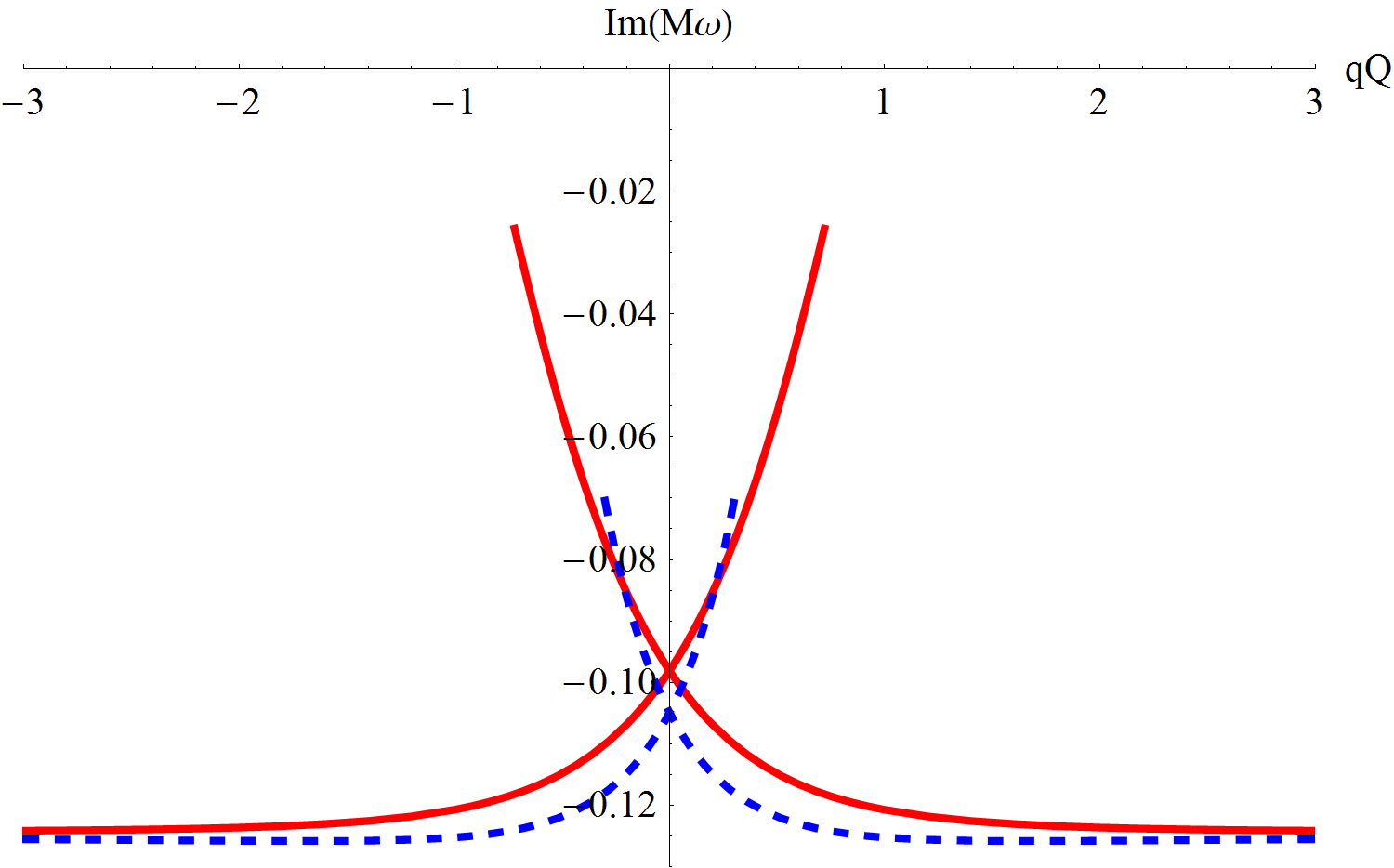}\\
\end{tabular}
\caption{Comparison between scalar ($s=j=0$, blue, dashed) and Dirac ($s=j=1/2$, red, solid) quasinormal modes. Real (left plot) and imaginary (right plot) parts are shown as functions of $qQ$ for $Q/M=0.5$. Note the critical values of the electromagnetic interaction around $|qQ|\approx 0.3$ for $s=j=0$ and around $|qQ| \approx 0.7$ for $s=j=1/2$.
\label{figc}
}
\end{center}
\end{figure*}

In Fig.~\ref{figb}, we have similar plots of the QNM frequencies as a function of $qQ$ for $Q/M=0.01$, $0.5$, $0.9$, $0.95$, $0.97$, $0.99$. However, due to the symmetry discussed above, we restrict the analysis to $Re(\omega)>0$ without loss of generality. Note the change in the behaviour of  $Im(\omega)$ as the extremal charge limit is approached, at some point between $Q/M=0.9$ and $Q/M=0.95$ (compare the top curves with the bottom ones). In all cases, as the electromagnetic interaction increases, the imaginary part of the quasinormal frequency approaches a constant value, while the real part seems to grow linearly. Besides that, the quasinormal frequencies for the scalar and fermionic cases approach the same value when $|qQ| \rightarrow \infty$, as one can observe in Fig.~\ref{figc}, where the scalar and Dirac spectra for $Q/M=0.5$ have been superimposed. These observations are also evident in table~\ref{table1} if one looks at it from top to bottom as $|qQ|$ is increased. Analytically, they can be confirmed by expanding the continued fraction equation~\eqref{cfeqn} in powers of $qQ$ and solving for $\omega$, producing 
\begin{align} \label{omexp}
\omega _ n = \frac{qQ}{r_+} - i \frac {(1+ 2n)(r_+-r_-)}{4r_+^2}     \nonumber \\
 + \frac{(r_+ - r_-) (2r_+(1 + 2s + 2 \lambda_s)+r_-(4s^2-1))}{16 r_+^3 qQ }  \nonumber \\
  - i \frac {(1+ 2n)(3r_--r_+)(4s^2 -1 )(r_+-r_-)^2}{64r_+^4 q^2Q^2}   ,
\end{align}
plus terms of order $(qQ)^{-3}$. A similar result [up to order  $(qQ)^{-1}$] was obtained previously in Refs.~\cite{hod} and~\cite{konoplya}, but only for the scalar case. We would like to note that our result also holds in the fermionic case, and that deviations from the scalar case only show up in terms of order $(qQ)^{-1}$ and higher, as shown explicitly in the equation above. 

\begin{figure*}
\begin{center}
\begin{tabular}{cc}
\includegraphics[width=8.6cm]{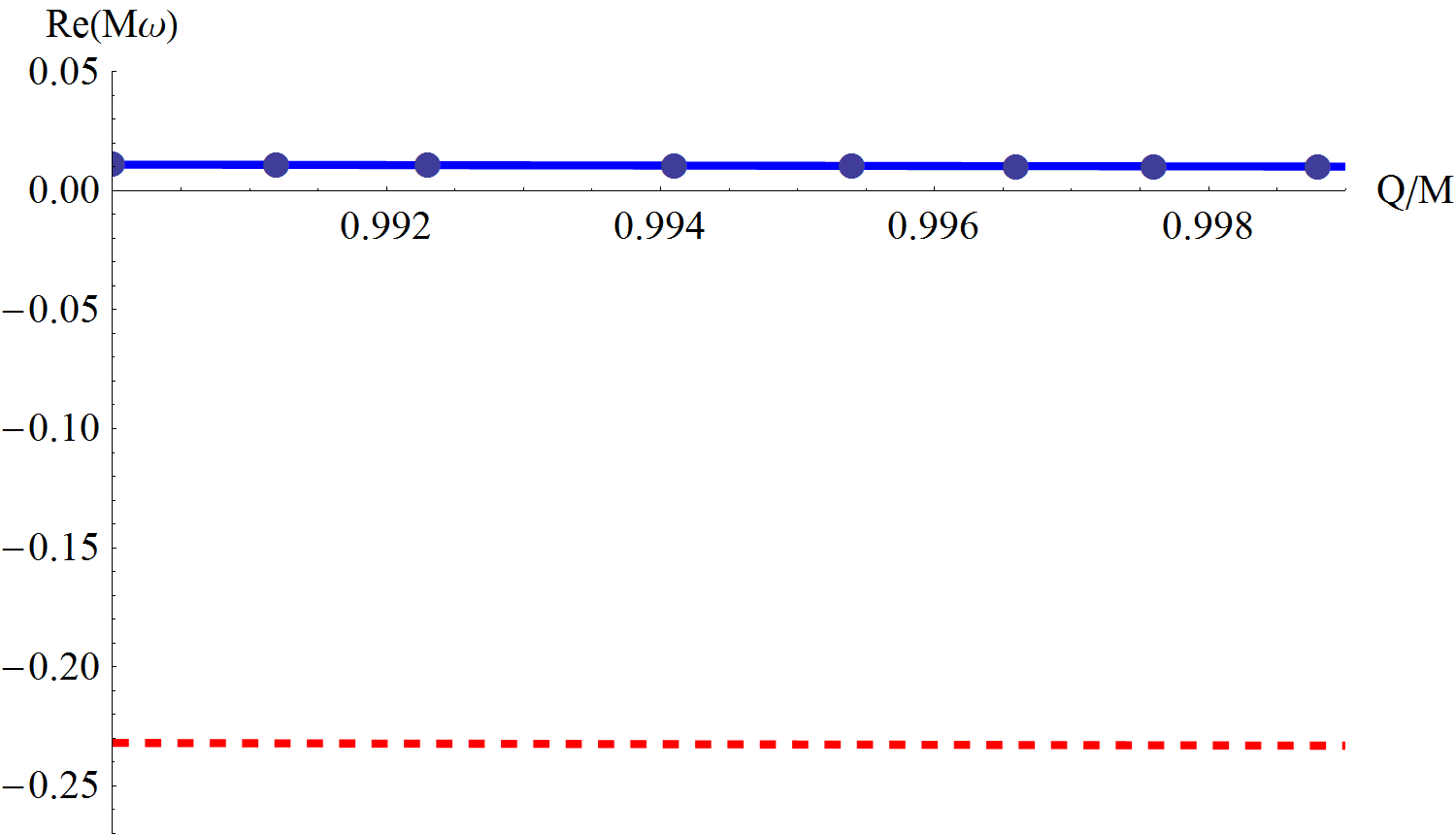}&
\includegraphics[width=8.6cm]{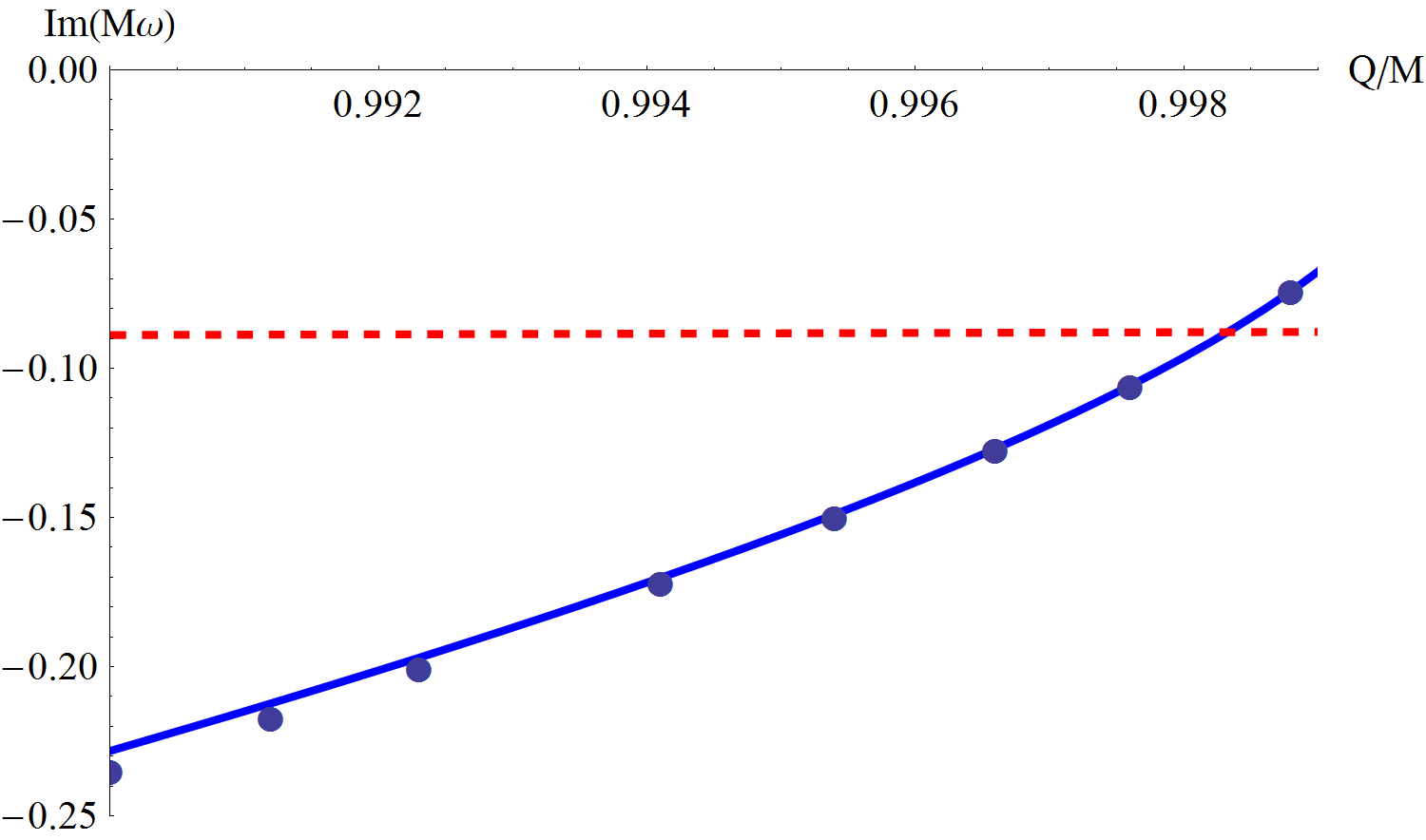}\\
\includegraphics[width=8.6cm]{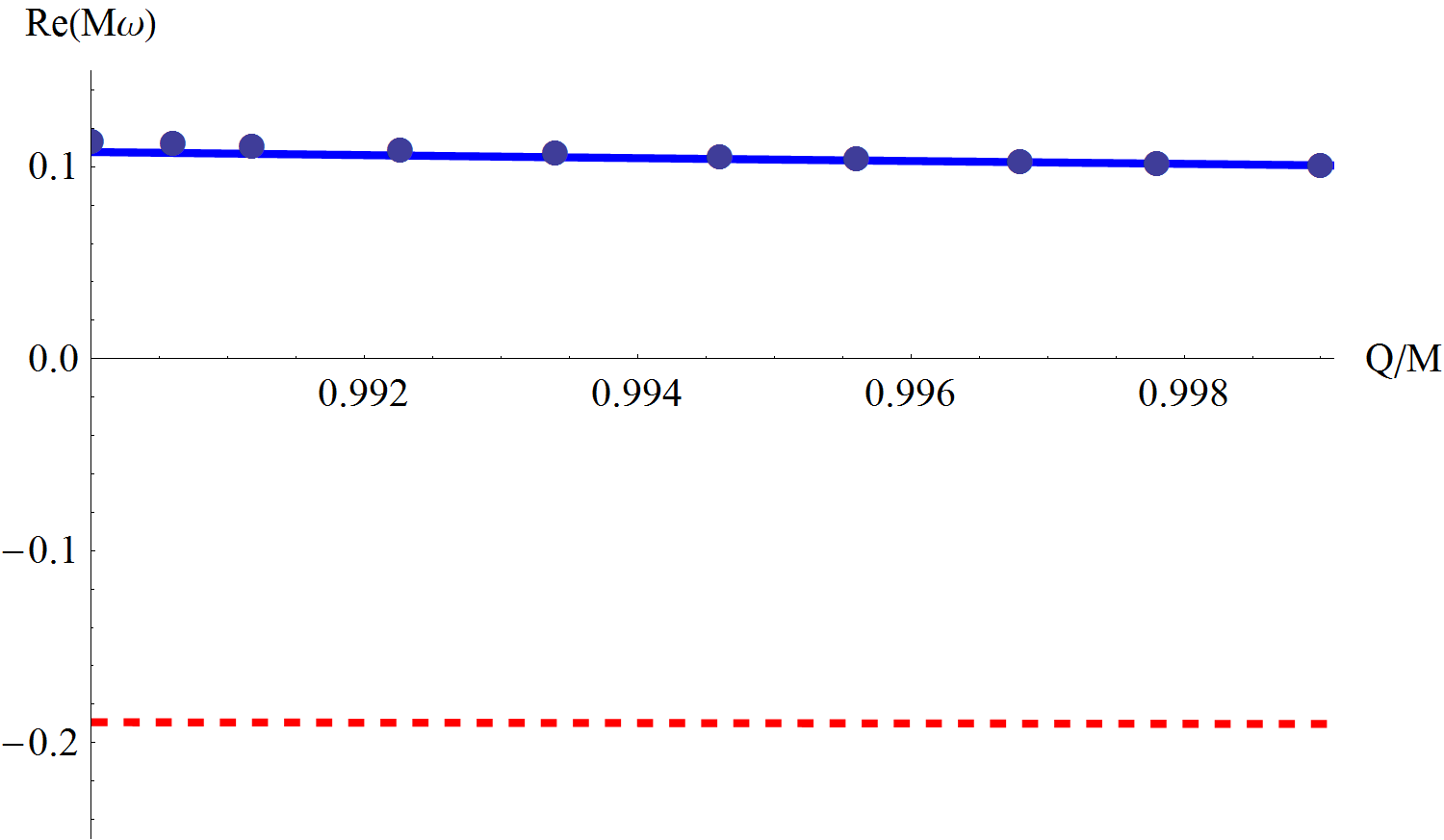}&
\includegraphics[width=8.6cm]{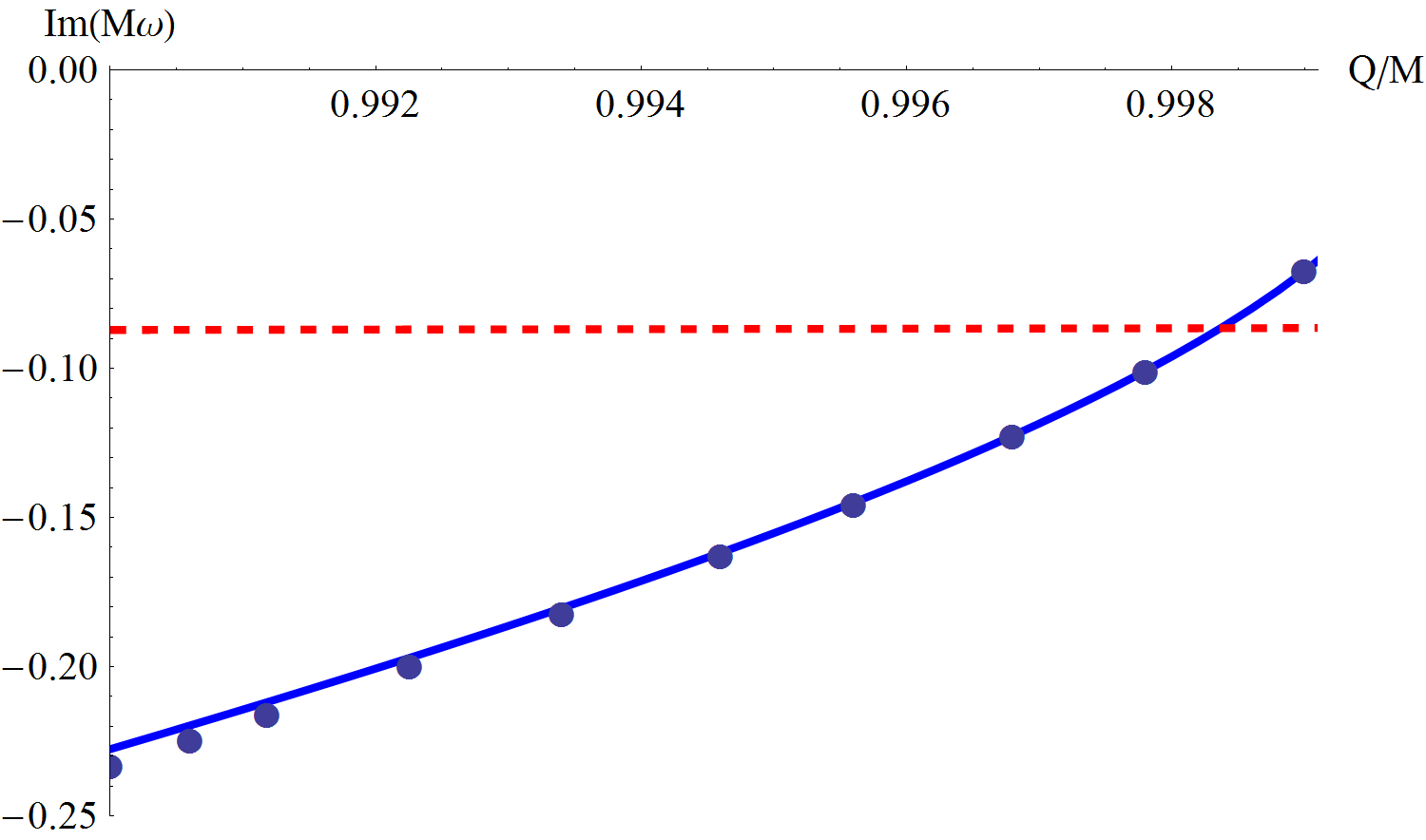}\\
\end{tabular}
\caption{Real (left plots) and imaginary (right plots) parts of the Dirac quasinormal modes ($s=j=1/2$) near the extremal limit for $qQ=0.01$ (top panels) and $qQ=0.1$ (bottom panels). Note that near the extremal limit (above $Q/M \approx 0.9985$), the most stable quasinormal mode corresponds to the blue-solid branch, with a real part given approximately by $Re(\omega)\approx qQ/r_+ \approx q$. Far from the extremal limit, the red-dashed branch corresponds to the fundamental frequency. The dots, which correspond to solutions of eq.~\eqref{algeb}, are in excellent agreement with the results of the continued fraction method.}
\label{figd}
\end{center}
\end{figure*}

Another important result which we have observed in our simulations (see the last column of table~\ref{table1}) is that the imaginary part of the fundamental frequency becomes smaller as the extremal limit $Q=M$ is approached, while the real part approaches $qQ/r_+ \approx q$. However, these modes have a strange behaviour since they correspond to solutions of the continued fraction equation which do not exist in the neutral limit $qQ=0$. More precisely, as pointed out in Ref.~\cite{konoplya} for the scalar case, if we fix $Q/M$ and take the limit $qQ\rightarrow 0$, there will be some critical value of $qQ$ below which these modes disappear from the spectrum. Similarly to the scalar case, these modes exist for all values of $Q/M$, but far from the extremal limit they correspond to higher overtones and therefore cannot be the fundamental mode (see Fig.~\ref{figd} for a plot of these `strange' modes showing the exact value of $Q/M$ above which they become fundamental modes).

Analytically, an indication of the existence of such modes can be obtained if, instead of solving eq.~\eqref{equat1} for $\omega$, we write $\omega = qQ/r_+ + K(r_+-r_-)$ and solve the corresponding equation for $K$. An interesting consequence is that the coefficients $\alpha_n$, $\beta_n$ and $\gamma_n$ become proportional to $(r_+-r_-)$. Eliminating this common factor, the coefficients can be redefined as
\begin{align}
\alpha_n =  -(1 + n)  (1 + n - 2 i K r_+^2 - s) ; \\
\beta_n  =1 - i qQ + 2 iKr_+ r_- + 
  2 (n - 2 i K r_+^2) \\ 
 [n +1 -iqQ -2iKr_+(r_+-r_-)] +   s + \lambda_s; \nonumber \\
\gamma_n = \frac{i}{r_+} (2 qQ r_- + i n r_+ - 2 K r_-^2 r_+ + 2 K r_+^3) \\ 
(n - 2 i K r_+^2 + s)  \nonumber.
\end{align}  
and then eq.~\eqref{equat1} can be normally solved for $K$. With the original coefficients, eq.~\eqref{equat1} is always proportional to $(r_+-r_-)$ and therefore one needs more precision in the root finding algorithm to correctly determine the QNMs when the extremal limit is approached.   

 The observations above suggest that, for exactly extremal black holes, the fundamental quasinormal frequencies for charged fields (scalar and Dirac) have vanishing imaginary parts, with real parts given by the field's charge $q$. An analogous result was derived by Detweiler~\cite{detweiler,cardosox} for nearly extreme Kerr black holes (through the calculation of the poles of the reflection coefficient), leading him to suggest that extremal Kerr black holes are marginally unstable. It was later argued that this marginal instability does not really occur~\cite{ferrari,glampe,bertix}. In our case, the existence of these modes with arbitrarily small imaginary parts could be an artifact of the complicate continued fraction equation~\eqref{equat1}, as suggested in Ref.~\cite{konoplya}. Fortunately, in the near extremal limit, eq.~\eqref{master} is amenable to analytic methods (see the appendix) and one is able to verify the existence of these modes by an alternative calculation. In fact, the QNMs obtained through the algebraic equation~\eqref{algeb} are in excellent agreement with the ones obtained through the continued fraction method in the limit $Q \rightarrow M$, as shown in Fig.~\ref{figd}. We have also verified that the last two columns of table~\ref{table1} are compatible with the solutions of eq.~\eqref{algeb}. In order to complement our analysis, it would be very interesting to check if these modes lead to a marginal instability of the extremal RN black hole or not. 
 
 Additionally, we would like to point out that a similar behaviour of purely real frequencies is also observed for massive fields. As reported in~\cite{qrn1,qrn2}, arbitrarily long living modes (the so-called quasi-resonant modes) can exist if the field mass has special values. If one changes the boundary condition at infinity to consider quasi-bound states instead of quasinormal modes, similar results can be observed. In fact, it has been shown that the frequencies of quasi-bound states of massive scalar~\cite{herdeiro1} and Proca~\cite{herdeiro2} fields have zero imaginary part in the limit $Q \rightarrow M$. The results of Ref.~\cite{herdeiro2} even indicate that this behaviour may occur not only for extremal black holes, but also in more generic situations.  
              
%================
\section{Final Remarks}
%================

In this work, we have implemented Leaver's continued fraction method, together with Nollert's improvement, to determine the charged Dirac quasinormal modes of a RN black hole. We were able to compute the quasinormal frequencies for arbitrary values of the interaction parameter $qQ$.  For $qQ=0$, our results exactly agree with the ones obtained in Ref.~\cite{jing2}. For small values of $|qQ|$ our results agree with the ones obtained in Refs.~\cite{chang,wu} through approximation methods (WKB and P\"oschl--Teller potential). As we increase the electromagnetic interaction, our simulations indicate the existence of a critical value of $|qQ|$ for which the fundamental quasinormal modes have arbitrarily small real oscillation frequencies. In the limit of large $|qQ|$, on the other hand, the QNMs where obtained both numerically and analytically for the first time, and the results where compared with the scalar case. Furthermore, our description unifies the treatment of charged QNMs for scalar and Dirac fields through the master equation~\eqref{master} and the continued fraction coefficients \eqref{al}--\eqref{be}.

 The case of nearly extremal black holes is particularly interesting due to the appearance of `strange' modes which do not exist for uncharged fields. The imaginary component $Im(\omega)$ of such modes approach zero as $Q \rightarrow M$, while  $Re(\omega)$ approaches $qQ/r_+\approx q$. This is exactly the upper limit of frequencies for which superradiant scattering occurs~\cite{bekenstein,mauricio3}. The existence of these modes indicate that QNMs of exactly extremal black holes have vanishing imaginary parts, similarly to what occurs for Kerr black holes. 
 
It is important to remark, however, that Leaver's original method fails to converge for extremal black holes. More precisely, when $M=Q$, the regular singularities at $r=r_-$ and at $r=r_+$ merge at $r=M$, becoming an irregular singularity. Therefore it is hopeless that a power series expansion around $r=r_+$ [like~\eqref{exp}] will have a non-zero radius of convergence. Onozawa et al~\cite{ozonawa} have proposed a modification of Leaver's method to deal with such type of equations and have successfully applied it to uncharged fields around an extremal RN black hole. The obtained results are in very good agreement with the ones obtained for nearly extremal black holes using Leaver's original method. As discussed by Leaver in Ref.~\cite{leaver2} for uncharged fields, this is not a surprise. Even though the convergence of the original method becomes worse as the extremal limit is approached, it can still be used with great accuracy quite close to the extremal limit. In our case, we have confirmed this fact by checking the results of the continued fraction method with an alternative calculation. Nonetheless, it would be interesting to implement the ideas of Ref.~\cite{ozonawa} to calculate the QNM spectrum of charged fields around exactly extremal black holes.

Finally, we would like to point out another interesting fact concerning our simulations: we have not observed any QNM with positive imaginary part, indicating that non--extremal RN black holes are stable under massless charged perturbations.     

\acknowledgments
The authors are grateful to Alberto Saa for enlightening discussions. M. R. was funded by the S\~ao Paulo Research
Foundation (FAPESP), Grant No. 2013/09357-9.

\begin{widetext}

\appendix*
\section{}

In the limit of extremal charge $Q \rightarrow M$, together with the assumption $\omega \rightarrow qQ/r_+$, it is possible to find the QNMs of charged fields around a RN black hole by calculating the poles of the reflection/transmission coefficients of the associated scattering problem. Such analysis has been performed in Ref.~\cite{hodap} for scalar fields. In what follows, we generalize the analysis of Ref.~\cite{hodap} to include not only the scalar case $s=0$ but also the Dirac case $s=1/2$. It is important to remark that such type of calculation is based on the seminal work of Press and Teukolsky~\cite{teuko} and has been used for the first time by Detweiler to determine the QNMs of a nearly extreme Kerr black hole~\cite{detweiler}.

Inspired by the notation of Ref.~\cite{hodap}, we use new dimensionless variables, namely 
\begin{align}
x=\frac{r-r_+}{r_+}, \quad \tau=\frac{r_+-r_-}{r_+}, \quad \hat{\omega} = \omega r_+,\quad
 k=2\hat{\omega}-qQ, \quad \varpi = \hat \omega - qQ,  
\end{align}
in order to rewrite eq.~\eqref{master} as
\begin{align} \label{master2}
x^2 (x+\tau)^ 2 R'' + (s+1)(2x+\tau)x(x+\tau)R' + 
\left[ \frac{K^ 2}{r_+^ 2} - i s (2x+\tau) \frac{K}{r_+} + x(x+\tau)\left(4is\hat \omega x+ 2isk - \lambda_s \right) \right]R=0,
\end{align}
where $K/r_+ = \hat \omega x^ 2 + k x + \varpi$, $R=R(x)$ is the radial function, and $'$ denotes the derivative with respect to $x$. We are interested in solving the equation above in the double limit $\tau \rightarrow 0$, $\varpi \rightarrow 0$. To do that, we first note that in the far region $x \gg max(\tau,\varpi)$, eq.~\eqref{master2} can be approximated by
\be 
x^2 R''+2(s+1)xR' + \left[\left(\hat \omega x + k\right)^2 - 2is\left(\hat \omega x + k\right) + \left(4is\hat \omega x+ 2isk - \lambda_s \right) \right]R=0.
\ee
Its most general solution can be written in terms of confluent hypergeometric functions,
\be \label{ape1}
R=C_1e^{-i\hat \omega x} x^{-\frac{1}{2}-2+i\delta} {}_1F_{1}\left(\frac{1}{2}-s+ik+i\delta,1+2i\delta,2i\hat\omega x\right) + C_2 (\delta \rightarrow -\delta),
\ee
where $C_1$ and $C_2$ are constants, $\delta ^ 2 = k^ 2 - \left(j+ \frac{1}{2} \right)^2$, and the notation $\delta \rightarrow -\delta$ means that the preceding term should be repeated with $-\delta$ replacing $\delta$.

 Near the horizon ($x \ll 1$), on the other hand, the radial equation can be approximated by
\begin{align}
x^2 (x+\tau)^ 2 R'' + (s+1)(2x+\tau)x(x+\tau)R' + 
\left[ \left(kx+\varpi\right)^ 2 - i s (2x+\tau) \left(kx+\varpi\right) + x(x+\tau)\left(2isk - \lambda_s \right) \right]R=0.
\end{align}
The only solution of the equation above which is compatible with the boundary condition of no outgoing waves near the event horizon is 
\be \label{ape2}
R = x^{-s - i \frac{\varpi}{\tau}} \left(x+\tau\right)^{-s - ik + i \frac{\varpi}{\tau}} {}_2F_{1}\left(\frac{1}{2}-s-ik+i\delta,\frac{1}{2}-s-ik-i\delta,1-s-2i\frac{\varpi}{\tau},-\frac{x}{\tau}\right).
\ee

By matching the solutions~\eqref{ape1} and \eqref{ape2} above in the overlap region $max(\tau,\varpi)\ll x \ll 1$, the coefficients $C_1$ and $C_2$ can be determined: 
\be
C_1=\frac{\Gamma\left(1-s-2i\frac{\varpi}{\tau}\right)\Gamma\left(2i\delta\right) \tau ^ {\frac{1}{2}-s-ik-i\delta}}{\Gamma\left(\frac{1}{2}-s-ik+i\delta\right)\Gamma\left(\frac{1}{2}+ik+i\delta-2i\frac{\varpi}{\tau}\right)}, \qquad
C_2=\frac{\Gamma\left(1-s-2i\frac{\varpi}{\tau}\right)\Gamma\left(-2i\delta\right)\tau ^ {\frac{1}{2}-s-ik+i\delta}}{\Gamma\left(\frac{1}{2}-s-ik-i\delta\right)\Gamma\left(\frac{1}{2}+ik-i\delta-2i\frac{\varpi}{\tau}\right)}. 
\ee

Finally, the requirement that eq.~\eqref{ape1} be compatible with \eqref{asymp} in the limit $x\rightarrow \infty$ (i.e. no ingoing waves) produces an algebraic equation for the QNM frequencies:
\be \label{algeb}
\frac{\Gamma\left(2i\delta\right)\Gamma\left(1+2i\delta\right)\Gamma\left(\frac{1}{2}-s-ik-i\delta\right)\Gamma\left(\frac{1}{2}+s-ik-i\delta\right)\Gamma\left(\frac{1}{2}+ik-i\delta-2i\frac{\varpi}{\tau}\right)}{\Gamma\left(-2i\delta\right)\Gamma\left(1-2i\delta\right)\Gamma\left(\frac{1}{2}-s-ik+i\delta\right)\Gamma\left(\frac{1}{2}+s-ik+i\delta\right)\Gamma\left(\frac{1}{2}+ik+i\delta-2i\frac{\varpi}{\tau}\right)}+\left(-2i\hat \omega \tau \right)^{2i\delta}=0
\ee

As expected, the equation above reduces to the expression obtained in Ref.~\cite{hodap} when $s=0$. We also note that a similar calculation was performed in Ref.~\cite{jing} for charged perturbations of a Kerr--Newman black hole. However, the results of Ref.~\cite{jing} do not seem to reduce to the expression above in the limit of zero rotation (and consequently, they are also incompatible with the results of Ref.~\cite{hodap}).  
\end{widetext}

%\bibliographystyle{apsrev4-1}
%\bibliography{qn_bib}

\begin{thebibliography}{49}%
\makeatletter
\providecommand \@ifxundefined [1]{%
 \@ifx{#1\undefined}
}%
\providecommand \@ifnum [1]{%
 \ifnum #1\expandafter \@firstoftwo
 \else \expandafter \@secondoftwo
 \fi
}%
\providecommand \@ifx [1]{%
 \ifx #1\expandafter \@firstoftwo
 \else \expandafter \@secondoftwo
 \fi
}%
\providecommand \natexlab [1]{#1}%
\providecommand \enquote  [1]{``#1''}%
\providecommand \bibnamefont  [1]{#1}%
\providecommand \bibfnamefont [1]{#1}%
\providecommand \citenamefont [1]{#1}%
\providecommand \href@noop [0]{\@secondoftwo}%
\providecommand \href [0]{\begingroup \@sanitize@url \@href}%
\providecommand \@href[1]{\@@startlink{#1}\@@href}%
\providecommand \@@href[1]{\endgroup#1\@@endlink}%
\providecommand \@sanitize@url [0]{\catcode `\\12\catcode `\$12\catcode
  `\&12\catcode `\#12\catcode `\^12\catcode `\_12\catcode `\%12\relax}%
\providecommand \@@startlink[1]{}%
\providecommand \@@endlink[0]{}%
\providecommand \url  [0]{\begingroup\@sanitize@url \@url }%
\providecommand \@url [1]{\endgroup\@href {#1}{\urlprefix }}%
\providecommand \urlprefix  [0]{URL }%
\providecommand \Eprint [0]{\href }%
\providecommand \doibase [0]{http://dx.doi.org/}%
\providecommand \selectlanguage [0]{\@gobble}%
\providecommand \bibinfo  [0]{\@secondoftwo}%
\providecommand \bibfield  [0]{\@secondoftwo}%
\providecommand \translation [1]{[#1]}%
\providecommand \BibitemOpen [0]{}%
\providecommand \bibitemStop [0]{}%
\providecommand \bibitemNoStop [0]{.\EOS\space}%
\providecommand \EOS [0]{\spacefactor3000\relax}%
\providecommand \BibitemShut  [1]{\csname bibitem#1\endcsname}%
\let\auto@bib@innerbib\@empty
%</preamble>
\bibitem [{\citenamefont {Regge}\ and\ \citenamefont {Wheeler}(1957)}]{regge}%
  \BibitemOpen
  \bibfield  {author} {\bibinfo {author} {\bibfnamefont {T.}~\bibnamefont
  {Regge}}\ and\ \bibinfo {author} {\bibfnamefont {J.~A.}\ \bibnamefont
  {Wheeler}},\ }\href {\doibase 10.1103/PhysRev.108.1063} {\bibfield  {journal}
  {\bibinfo  {journal} {Phys.Rev.}\ }\textbf {\bibinfo {volume} {108}},\
  \bibinfo {pages} {1063} (\bibinfo {year} {1957})}\BibitemShut {NoStop}%
%%CITATION = PHRVA,108,1063;%%
\bibitem [{\citenamefont {Vishveshwara}(1970)}]{vish}%
  \BibitemOpen
  \bibfield  {author} {\bibinfo {author} {\bibfnamefont {C.}~\bibnamefont
  {Vishveshwara}},\ }\href {\doibase 10.1038/227936a0} {\bibfield  {journal}
  {\bibinfo  {journal} {Nature}\ }\textbf {\bibinfo {volume} {227}},\ \bibinfo
  {pages} {936} (\bibinfo {year} {1970})}\BibitemShut {NoStop}%
%%CITATION = NATUA,227,936;%%
\bibitem [{\citenamefont {Press}(1971)}]{press}%
  \BibitemOpen
  \bibfield  {author} {\bibinfo {author} {\bibfnamefont {W.~H.}\ \bibnamefont
  {Press}},\ }\href {\doibase 10.1086/180849} {\bibfield  {journal} {\bibinfo
  {journal} {Astrophys.J.}\ }\textbf {\bibinfo {volume} {170}},\ \bibinfo
  {pages} {L105} (\bibinfo {year} {1971})}\BibitemShut {NoStop}%
%%CITATION = ASJOA,170,L105;%%
\bibitem [{\citenamefont {Detweiler}(1977)}]{det1}%
  \BibitemOpen
  \bibfield  {author} {\bibinfo {author} {\bibfnamefont {S.~L.}\ \bibnamefont
  {Detweiler}},\ }\href {\doibase 10.1098/rspa.1977.0005} {\bibfield  {journal}
  {\bibinfo  {journal} {Proc.Roy.Soc.Lond.}\ }\textbf {\bibinfo {volume}
  {A352}},\ \bibinfo {pages} {381} (\bibinfo {year} {1977})}\BibitemShut
  {NoStop}%
%%CITATION = PRSLA,A352,381;%%
\bibitem [{\citenamefont {Detweiler}\ and\ \citenamefont
  {Szedenits}(1979)}]{det2}%
  \BibitemOpen
  \bibfield  {author} {\bibinfo {author} {\bibfnamefont {S.~L.}\ \bibnamefont
  {Detweiler}}\ and\ \bibinfo {author} {\bibfnamefont {E.}~\bibnamefont
  {Szedenits}},\ }\href {\doibase 10.1086/157182} {\bibfield  {journal}
  {\bibinfo  {journal} {Astrophys.J.}\ }\textbf {\bibinfo {volume} {231}},\
  \bibinfo {pages} {211} (\bibinfo {year} {1979})}\BibitemShut {NoStop}%
%%CITATION = ASJOA,231,211;%%
\bibitem [{\citenamefont {Davis}\ \emph {et~al.}(1971)\citenamefont {Davis},
  \citenamefont {Ruffini}, \citenamefont {Press},\ and\ \citenamefont
  {Price}}]{davis}%
  \BibitemOpen
  \bibfield  {author} {\bibinfo {author} {\bibfnamefont {M.}~\bibnamefont
  {Davis}}, \bibinfo {author} {\bibfnamefont {R.}~\bibnamefont {Ruffini}},
  \bibinfo {author} {\bibfnamefont {W.}~\bibnamefont {Press}}, \ and\ \bibinfo
  {author} {\bibfnamefont {R.}~\bibnamefont {Price}},\ }\href {\doibase
  10.1103/PhysRevLett.27.1466} {\bibfield  {journal} {\bibinfo  {journal}
  {Phys.Rev.Lett.}\ }\textbf {\bibinfo {volume} {27}},\ \bibinfo {pages} {1466}
  (\bibinfo {year} {1971})}\BibitemShut {NoStop}%
%%CITATION = PRLTA,27,1466;%%
\bibitem [{\citenamefont {Cunningham}\ \emph {et~al.}(1978)\citenamefont
  {Cunningham}, \citenamefont {Price},\ and\ \citenamefont {Moncrief}}]{cunn1}%
  \BibitemOpen
  \bibfield  {author} {\bibinfo {author} {\bibfnamefont {C.}~\bibnamefont
  {Cunningham}}, \bibinfo {author} {\bibfnamefont {R.}~\bibnamefont {Price}}, \
  and\ \bibinfo {author} {\bibfnamefont {V.}~\bibnamefont {Moncrief}},\ }\href
  {\doibase 10.1086/156413} {\bibfield  {journal} {\bibinfo  {journal}
  {Astrophys.J.}\ }\textbf {\bibinfo {volume} {224}},\ \bibinfo {pages} {643}
  (\bibinfo {year} {1978})}\BibitemShut {NoStop}%
%%CITATION = ASJOA,224,643;%%
\bibitem [{\citenamefont {Cunningham}\ \emph {et~al.}(1979)\citenamefont
  {Cunningham}, \citenamefont {Price},\ and\ \citenamefont {Moncrief}}]{cunn2}%
  \BibitemOpen
  \bibfield  {author} {\bibinfo {author} {\bibfnamefont {C.}~\bibnamefont
  {Cunningham}}, \bibinfo {author} {\bibfnamefont {R.}~\bibnamefont {Price}}, \
  and\ \bibinfo {author} {\bibfnamefont {V.}~\bibnamefont {Moncrief}},\ }\href
  {\doibase 10.1086/157147} {\bibfield  {journal} {\bibinfo  {journal}
  {Astrophys.J.}\ }\textbf {\bibinfo {volume} {230}},\ \bibinfo {pages} {870}
  (\bibinfo {year} {1979})}\BibitemShut {NoStop}%
%%CITATION = ASJOA,230,870;%%
\bibitem [{\citenamefont {{Press}}\ and\ \citenamefont
  {{Teukolsky}}(1973)}]{teuko}%
  \BibitemOpen
  \bibfield  {author} {\bibinfo {author} {\bibfnamefont {W.~H.}\ \bibnamefont
  {{Press}}}\ and\ \bibinfo {author} {\bibfnamefont {S.~A.}\ \bibnamefont
  {{Teukolsky}}},\ }\href {\doibase 10.1086/152445} {\bibfield  {journal}
  {\bibinfo  {journal} {\apj}\ }\textbf {\bibinfo {volume} {185}},\ \bibinfo
  {pages} {649} (\bibinfo {year} {1973})}\BibitemShut {NoStop}%
\bibitem [{\citenamefont {Whiting}(1989)}]{whiting}%
  \BibitemOpen
  \bibfield  {author} {\bibinfo {author} {\bibfnamefont {B.~F.}\ \bibnamefont
  {Whiting}},\ }\href {\doibase 10.1063/1.528308} {\bibfield  {journal}
  {\bibinfo  {journal} {J. Math. Phys.}\ }\textbf {\bibinfo {volume} {30}},\
  \bibinfo {pages} {1301} (\bibinfo {year} {1989})}\BibitemShut {NoStop}%
\bibitem [{\citenamefont {{Hod}}(1998)}]{hodquantum}%
  \BibitemOpen
  \bibfield  {author} {\bibinfo {author} {\bibfnamefont {S.}~\bibnamefont
  {{Hod}}},\ }\href {\doibase 10.1103/PhysRevLett.81.4293} {\bibfield
  {journal} {\bibinfo  {journal} {Physical Review Letters}\ }\textbf {\bibinfo
  {volume} {81}},\ \bibinfo {pages} {4293} (\bibinfo {year}
  {1998})}\BibitemShut {NoStop}%
\bibitem [{\citenamefont {{Hod}}(2006)}]{hodquantum2}%
  \BibitemOpen
  \bibfield  {author} {\bibinfo {author} {\bibfnamefont {S.}~\bibnamefont
  {{Hod}}},\ }\href {\doibase 10.1088/0264-9381/23/4/L01} {\bibfield  {journal}
  {\bibinfo  {journal} {Classical and Quantum Gravity}\ }\textbf {\bibinfo
  {volume} {23}},\ \bibinfo {pages} {L23} (\bibinfo {year} {2006})}\BibitemShut
  {NoStop}%
\bibitem [{\citenamefont {{Corda}}(2012)}]{corda}%
  \BibitemOpen
  \bibfield  {author} {\bibinfo {author} {\bibfnamefont {C.}~\bibnamefont
  {{Corda}}},\ }\href {\doibase 10.1142/S0218271812420230} {\bibfield
  {journal} {\bibinfo  {journal} {Int.J.Mod.Phys. D}\ }\textbf {\bibinfo
  {volume} {21}},\ \bibinfo {eid} {1242023} (\bibinfo {year}
  {2012})}\BibitemShut {NoStop}%
\bibitem [{\citenamefont {{Birmingham}}\ \emph {et~al.}(2002)\citenamefont
  {{Birmingham}}, \citenamefont {{Sachs}},\ and\ \citenamefont
  {{Solodukhin}}}]{ads}%
  \BibitemOpen
  \bibfield  {author} {\bibinfo {author} {\bibfnamefont {D.}~\bibnamefont
  {{Birmingham}}}, \bibinfo {author} {\bibfnamefont {I.}~\bibnamefont
  {{Sachs}}}, \ and\ \bibinfo {author} {\bibfnamefont {S.~N.}\ \bibnamefont
  {{Solodukhin}}},\ }\href@noop {} {\bibfield  {journal} {\bibinfo  {journal}
  {Physical Review Letters}\ }\textbf {\bibinfo {volume} {88}},\ \bibinfo {eid}
  {151301} (\bibinfo {year} {2002})}\BibitemShut {NoStop}%
\bibitem [{\citenamefont {Berti}\ \emph {et~al.}(2004)\citenamefont {Berti},
  \citenamefont {Cardoso},\ and\ \citenamefont {Lemos}}]{ana}%
  \BibitemOpen
  \bibfield  {author} {\bibinfo {author} {\bibfnamefont {E.}~\bibnamefont
  {Berti}}, \bibinfo {author} {\bibfnamefont {V.}~\bibnamefont {Cardoso}}, \
  and\ \bibinfo {author} {\bibfnamefont {J.~P.~S.}\ \bibnamefont {Lemos}},\
  }\href@noop {} {\bibfield  {journal} {\bibinfo  {journal} {Phys. Rev. D}\
  }\textbf {\bibinfo {volume} {70}},\ \bibinfo {eid} {124006} (\bibinfo {year}
  {2004})}\BibitemShut {NoStop}%
\bibitem [{\citenamefont {Kokkotas}\ and\ \citenamefont
  {Schmidt}(1999)}]{review1}%
  \BibitemOpen
  \bibfield  {author} {\bibinfo {author} {\bibfnamefont {K.~D.}\ \bibnamefont
  {Kokkotas}}\ and\ \bibinfo {author} {\bibfnamefont {B.~G.}\ \bibnamefont
  {Schmidt}},\ }\href@noop {} {\bibfield  {journal} {\bibinfo  {journal}
  {Living Rev.Rel.}\ }\textbf {\bibinfo {volume} {2}},\ \bibinfo {pages} {2}
  (\bibinfo {year} {1999})}\BibitemShut {NoStop}%
\bibitem [{\citenamefont {Nollert}(1999)}]{review2}%
  \BibitemOpen
  \bibfield  {author} {\bibinfo {author} {\bibfnamefont {H.-P.}\ \bibnamefont
  {Nollert}},\ }\href {\doibase 10.1088/0264-9381/16/12/201} {\bibfield
  {journal} {\bibinfo  {journal} {Class.Quant.Grav.}\ }\textbf {\bibinfo
  {volume} {16}},\ \bibinfo {pages} {R159} (\bibinfo {year}
  {1999})}\BibitemShut {NoStop}%
%%CITATION = CQGRD,16,R159;%%
\bibitem [{\citenamefont {Berti}\ \emph {et~al.}(2009)\citenamefont {Berti},
  \citenamefont {Cardoso},\ and\ \citenamefont {Starinets}}]{review3}%
  \BibitemOpen
  \bibfield  {author} {\bibinfo {author} {\bibfnamefont {E.}~\bibnamefont
  {Berti}}, \bibinfo {author} {\bibfnamefont {V.}~\bibnamefont {Cardoso}}, \
  and\ \bibinfo {author} {\bibfnamefont {A.~O.}\ \bibnamefont {Starinets}},\
  }\href@noop {} {\bibfield  {journal} {\bibinfo  {journal}
  {Class.Quant.Grav.}\ }\textbf {\bibinfo {volume} {26}},\ \bibinfo {pages}
  {163001} (\bibinfo {year} {2009})}\BibitemShut {NoStop}%
\bibitem [{\citenamefont {{Konoplya}}\ and\ \citenamefont
  {{Zhidenko}}(2011)}]{review4}%
  \BibitemOpen
  \bibfield  {author} {\bibinfo {author} {\bibfnamefont {R.~A.}\ \bibnamefont
  {{Konoplya}}}\ and\ \bibinfo {author} {\bibfnamefont {A.}~\bibnamefont
  {{Zhidenko}}},\ }\href@noop {} {\bibfield  {journal} {\bibinfo  {journal}
  {Reviews of Modern Physics}\ }\textbf {\bibinfo {volume} {83}},\ \bibinfo
  {pages} {793} (\bibinfo {year} {2011})}\BibitemShut {NoStop}%
\bibitem [{\citenamefont {Konoplya}(2002{\natexlab{a}})}]{konoa}%
  \BibitemOpen
  \bibfield  {author} {\bibinfo {author} {\bibfnamefont {R.}~\bibnamefont
  {Konoplya}},\ }\href {\doibase 10.1016/S0370-2693(02)02974-X} {\bibfield
  {journal} {\bibinfo  {journal} {Phys.Lett.}\ }\textbf {\bibinfo {volume}
  {B550}},\ \bibinfo {pages} {117} (\bibinfo {year}
  {2002}{\natexlab{a}})}\BibitemShut {NoStop}%
\bibitem [{\citenamefont {Konoplya}(2002{\natexlab{b}})}]{konob}%
  \BibitemOpen
  \bibfield  {author} {\bibinfo {author} {\bibfnamefont {R.}~\bibnamefont
  {Konoplya}},\ }\href {\doibase 10.1103/PhysRevD.66.084007} {\bibfield
  {journal} {\bibinfo  {journal} {Phys.Rev.}\ }\textbf {\bibinfo {volume}
  {D66}},\ \bibinfo {pages} {084007} (\bibinfo {year}
  {2002}{\natexlab{b}})}\BibitemShut {NoStop}%
\bibitem [{\citenamefont {{Konoplya}}\ and\ \citenamefont
  {{Zhidenko}}(2007)}]{zhid}%
  \BibitemOpen
  \bibfield  {author} {\bibinfo {author} {\bibfnamefont {R.~A.}\ \bibnamefont
  {{Konoplya}}}\ and\ \bibinfo {author} {\bibfnamefont {A.}~\bibnamefont
  {{Zhidenko}}},\ }\href {\doibase 10.1103/PhysRevD.76.084018} {\bibfield
  {journal} {\bibinfo  {journal} {\prd}\ }\textbf {\bibinfo {volume} {76}},\
  \bibinfo {eid} {084018} (\bibinfo {year} {2007})}\BibitemShut {NoStop}%
\bibitem [{\citenamefont {Chang}\ and\ \citenamefont {Shen}(2007)}]{chang}%
  \BibitemOpen
  \bibfield  {author} {\bibinfo {author} {\bibfnamefont {J.}~\bibnamefont
  {Chang}}\ and\ \bibinfo {author} {\bibfnamefont {Y.}~\bibnamefont {Shen}},\
  }\href@noop {} {\bibfield  {journal} {\bibinfo  {journal} {International
  Journal of Theoretical Physics}\ }\textbf {\bibinfo {volume} {46}},\ \bibinfo
  {pages} {1570} (\bibinfo {year} {2007})}\BibitemShut {NoStop}%
\bibitem [{\citenamefont {Jing}\ \emph {et~al.}(2007)\citenamefont {Jing},
  \citenamefont {Pan},\ and\ \citenamefont {He}}]{jing}%
  \BibitemOpen
  \bibfield  {author} {\bibinfo {author} {\bibfnamefont {J.}~\bibnamefont
  {Jing}}, \bibinfo {author} {\bibfnamefont {Q.-Y.}\ \bibnamefont {Pan}}, \
  and\ \bibinfo {author} {\bibfnamefont {X.}~\bibnamefont {He}},\ }\href
  {\doibase 10.1142/S0218271807009322} {\bibfield  {journal} {\bibinfo
  {journal} {Int.J.Mod.Phys. D}\ }\textbf {\bibinfo {volume} {D16}},\ \bibinfo
  {pages} {81} (\bibinfo {year} {2007})}\BibitemShut {NoStop}%
%%CITATION = IMPAE,D16,81;%%
\bibitem [{\citenamefont {Wu}\ and\ \citenamefont {Zhao}(2004)}]{wu}%
  \BibitemOpen
  \bibfield  {author} {\bibinfo {author} {\bibfnamefont {Y.-J.}\ \bibnamefont
  {Wu}}\ and\ \bibinfo {author} {\bibfnamefont {Z.}~\bibnamefont {Zhao}},\
  }\href@noop {} {\bibfield  {journal} {\bibinfo  {journal} {Phys. Rev. D}\
  }\textbf {\bibinfo {volume} {69}},\ \bibinfo {pages} {084015} (\bibinfo
  {year} {2004})}\BibitemShut {NoStop}%
\bibitem [{\citenamefont {{Hod}}(2010)}]{hodap}%
  \BibitemOpen
  \bibfield  {author} {\bibinfo {author} {\bibfnamefont {S.}~\bibnamefont
  {{Hod}}},\ }\href {\doibase 10.1016/j.physleta.2010.05.052} {\bibfield
  {journal} {\bibinfo  {journal} {Physics Letters A}\ }\textbf {\bibinfo
  {volume} {374}},\ \bibinfo {pages} {2901} (\bibinfo {year}
  {2010})}\BibitemShut {NoStop}%
\bibitem [{\citenamefont {Hod}(2012)}]{hod}%
  \BibitemOpen
  \bibfield  {author} {\bibinfo {author} {\bibfnamefont {S.}~\bibnamefont
  {Hod}},\ }\href@noop {} {\bibfield  {journal} {\bibinfo  {journal}
  {Phys.Lett.}\ }\textbf {\bibinfo {volume} {B710}},\ \bibinfo {pages} {349}
  (\bibinfo {year} {2012})}\BibitemShut {NoStop}%
\bibitem [{\citenamefont {Konoplya}\ and\ \citenamefont
  {Zhidenko}(2013)}]{konoplya}%
  \BibitemOpen
  \bibfield  {author} {\bibinfo {author} {\bibfnamefont {R.}~\bibnamefont
  {Konoplya}}\ and\ \bibinfo {author} {\bibfnamefont {A.}~\bibnamefont
  {Zhidenko}},\ }\href@noop {} {\bibfield  {journal} {\bibinfo  {journal}
  {Phys.Rev.}\ }\textbf {\bibinfo {volume} {D88}},\ \bibinfo {pages} {024054}
  (\bibinfo {year} {2013})}\BibitemShut {NoStop}%
\bibitem [{\citenamefont {Stewart}(1991)}]{stewart}%
  \BibitemOpen
  \bibfield  {author} {\bibinfo {author} {\bibfnamefont {J.}~\bibnamefont
  {Stewart}},\ }\href@noop {} {\emph {\bibinfo {title} {Advanced general
  relativity}}}\ (\bibinfo  {publisher} {Cambridge University Press},\ \bibinfo
  {year} {1991})\BibitemShut {NoStop}%
\bibitem [{\citenamefont {Goldberg}\ \emph {et~al.}(1967)\citenamefont
  {Goldberg}, \citenamefont {Macfarlane}, \citenamefont {Newman}, \citenamefont
  {Rohrlich},\ and\ \citenamefont {Sudarshan}}]{goldberg}%
  \BibitemOpen
  \bibfield  {author} {\bibinfo {author} {\bibfnamefont {J.~N.}\ \bibnamefont
  {Goldberg}}, \bibinfo {author} {\bibfnamefont {A.~J.}\ \bibnamefont
  {Macfarlane}}, \bibinfo {author} {\bibfnamefont {E.~T.}\ \bibnamefont
  {Newman}}, \bibinfo {author} {\bibfnamefont {F.}~\bibnamefont {Rohrlich}}, \
  and\ \bibinfo {author} {\bibfnamefont {E.~C.~G.}\ \bibnamefont {Sudarshan}},\
  }\href@noop {} {\bibfield  {journal} {\bibinfo  {journal} {J. Math. Phys.}\
  }\textbf {\bibinfo {volume} {8}},\ \bibinfo {pages} {2155} (\bibinfo {year}
  {1967})}\BibitemShut {NoStop}%
\bibitem [{\citenamefont {{Richartz}}\ and\ \citenamefont
  {{Saa}}(2011)}]{mauricio_alberto}%
  \BibitemOpen
  \bibfield  {author} {\bibinfo {author} {\bibfnamefont {M.}~\bibnamefont
  {{Richartz}}}\ and\ \bibinfo {author} {\bibfnamefont {A.}~\bibnamefont
  {{Saa}}},\ }\href {\doibase 10.1103/PhysRevD.84.104021} {\bibfield  {journal}
  {\bibinfo  {journal} {\prd}\ }\textbf {\bibinfo {volume} {84}},\ \bibinfo
  {eid} {104021} (\bibinfo {year} {2011})}\BibitemShut {NoStop}%
\bibitem [{\citenamefont {Teukolsky}(1972)}]{teukolsky}%
  \BibitemOpen
  \bibfield  {author} {\bibinfo {author} {\bibfnamefont {S.~A.}\ \bibnamefont
  {Teukolsky}},\ }\href@noop {} {\bibfield  {journal} {\bibinfo  {journal}
  {Phys. Rev. Lett.}\ }\textbf {\bibinfo {volume} {29}},\ \bibinfo {pages}
  {1114} (\bibinfo {year} {1972})}\BibitemShut {NoStop}%
\bibitem [{\citenamefont {Leaver}(1985)}]{leaver1}%
  \BibitemOpen
  \bibfield  {author} {\bibinfo {author} {\bibfnamefont {E.}~\bibnamefont
  {Leaver}},\ }\href {\doibase 10.1098/rspa.1985.0119} {\bibfield  {journal}
  {\bibinfo  {journal} {Proc.Roy.Soc.Lond.}\ }\textbf {\bibinfo {volume}
  {A402}},\ \bibinfo {pages} {285} (\bibinfo {year} {1985})}\BibitemShut
  {NoStop}%
%%CITATION = PRSLA,A402,285;%%
\bibitem [{\citenamefont {Leaver}(1990)}]{leaver2}%
  \BibitemOpen
  \bibfield  {author} {\bibinfo {author} {\bibfnamefont {E.~W.}\ \bibnamefont
  {Leaver}},\ }\href {\doibase 10.1103/PhysRevD.41.2986} {\bibfield  {journal}
  {\bibinfo  {journal} {Phys.Rev.}\ }\textbf {\bibinfo {volume} {D41}},\
  \bibinfo {pages} {2986} (\bibinfo {year} {1990})}\BibitemShut {NoStop}%
%%CITATION = PHRVA,D41,2986;%%
\bibitem [{\citenamefont {Nollert}(1993)}]{nollert}%
  \BibitemOpen
  \bibfield  {author} {\bibinfo {author} {\bibfnamefont {H.-P.}\ \bibnamefont
  {Nollert}},\ }\href {\doibase 10.1103/PhysRevD.47.5253} {\bibfield  {journal}
  {\bibinfo  {journal} {Phys.Rev.}\ }\textbf {\bibinfo {volume} {D47}},\
  \bibinfo {pages} {5253} (\bibinfo {year} {1993})}\BibitemShut {NoStop}%
%%CITATION = PHRVA,D47,5253;%%
\bibitem [{\citenamefont {Jing}(2005)}]{jing2}%
  \BibitemOpen
  \bibfield  {author} {\bibinfo {author} {\bibfnamefont {J.-l.}\ \bibnamefont
  {Jing}},\ }\href {\doibase 10.1088/1126-6708/2005/12/005} {\bibfield
  {journal} {\bibinfo  {journal} {JHEP}\ }\textbf {\bibinfo {volume} {0512}},\
  \bibinfo {pages} {005} (\bibinfo {year} {2005})}\BibitemShut {NoStop}%
\bibitem [{\citenamefont {Gautschi}(1967)}]{gaut}%
  \BibitemOpen
  \bibfield  {author} {\bibinfo {author} {\bibfnamefont {W.}~\bibnamefont
  {Gautschi}},\ }\href@noop {} {\bibfield  {journal} {\bibinfo  {journal} {SIAM
  Review}\ }\textbf {\bibinfo {volume} {9}},\ \bibinfo {pages} {24} (\bibinfo
  {year} {1967})}\BibitemShut {NoStop}%
\bibitem [{\citenamefont {{Detweiler}}(1980)}]{detweiler}%
  \BibitemOpen
  \bibfield  {author} {\bibinfo {author} {\bibfnamefont {S.}~\bibnamefont
  {{Detweiler}}},\ }\href@noop {} {\bibfield  {journal} {\bibinfo  {journal}
  {\apj}\ }\textbf {\bibinfo {volume} {239}},\ \bibinfo {pages} {292} (\bibinfo
  {year} {1980})}\BibitemShut {NoStop}%
\bibitem [{\citenamefont {Cardoso}(2004)}]{cardosox}%
  \BibitemOpen
  \bibfield  {author} {\bibinfo {author} {\bibfnamefont {V.}~\bibnamefont
  {Cardoso}},\ }\href {\doibase 10.1103/PhysRevD.70.127502} {\bibfield
  {journal} {\bibinfo  {journal} {Phys.Rev.}\ }\textbf {\bibinfo {volume}
  {D70}},\ \bibinfo {pages} {127502} (\bibinfo {year} {2004})}\BibitemShut
  {NoStop}%
\bibitem [{\citenamefont {Ferrari}\ and\ \citenamefont
  {Mashhoon}(1984)}]{ferrari}%
  \BibitemOpen
  \bibfield  {author} {\bibinfo {author} {\bibfnamefont {V.}~\bibnamefont
  {Ferrari}}\ and\ \bibinfo {author} {\bibfnamefont {B.}~\bibnamefont
  {Mashhoon}},\ }\href {\doibase 10.1103/PhysRevD.30.295} {\bibfield  {journal}
  {\bibinfo  {journal} {Phys. Rev. D}\ }\textbf {\bibinfo {volume} {30}},\
  \bibinfo {pages} {295} (\bibinfo {year} {1984})}\BibitemShut {NoStop}%
\bibitem [{\citenamefont {Glampedakis}\ and\ \citenamefont
  {Andersson}(2001)}]{glampe}%
  \BibitemOpen
  \bibfield  {author} {\bibinfo {author} {\bibfnamefont {K.}~\bibnamefont
  {Glampedakis}}\ and\ \bibinfo {author} {\bibfnamefont {N.}~\bibnamefont
  {Andersson}},\ }\href {\doibase 10.1103/PhysRevD.64.104021} {\bibfield
  {journal} {\bibinfo  {journal} {Phys.Rev.}\ }\textbf {\bibinfo {volume}
  {D64}},\ \bibinfo {pages} {104021} (\bibinfo {year} {2001})}\BibitemShut
  {NoStop}%
\bibitem [{\citenamefont {Berti}(2004)}]{bertix}%
  \BibitemOpen
  \bibfield  {author} {\bibinfo {author} {\bibfnamefont {E.}~\bibnamefont
  {Berti}},\ }\href@noop {} {\bibfield  {journal} {\bibinfo  {journal}
  {Conf.Proc.}\ }\textbf {\bibinfo {volume} {C0405132}},\ \bibinfo {pages}
  {145} (\bibinfo {year} {2004})}\BibitemShut {NoStop}%
\bibitem [{\citenamefont {Ohashi}\ and\ \citenamefont {Sakagami}(2004)}]{qrn1}%
  \BibitemOpen
  \bibfield  {author} {\bibinfo {author} {\bibfnamefont {A.}~\bibnamefont
  {Ohashi}}\ and\ \bibinfo {author} {\bibfnamefont {M.}~\bibnamefont
  {Sakagami}},\ }\href@noop {} {\bibfield  {journal} {\bibinfo  {journal}
  {Classical and Quantum Gravity}\ }\textbf {\bibinfo {volume} {21}},\ \bibinfo
  {pages} {3973} (\bibinfo {year} {2004})}\BibitemShut {NoStop}%
\bibitem [{\citenamefont {Konoplya}\ and\ \citenamefont
  {Zhidenko}(2005)}]{qrn2}%
  \BibitemOpen
  \bibfield  {author} {\bibinfo {author} {\bibfnamefont {R.}~\bibnamefont
  {Konoplya}}\ and\ \bibinfo {author} {\bibfnamefont {A.}~\bibnamefont
  {Zhidenko}},\ }\href@noop {} {\bibfield  {journal} {\bibinfo  {journal}
  {Phys.Lett.}\ }\textbf {\bibinfo {volume} {B609}},\ \bibinfo {pages} {377}
  (\bibinfo {year} {2005})}\BibitemShut {NoStop}%
\bibitem [{\citenamefont {{Degollado}}\ and\ \citenamefont
  {{Herdeiro}}(2013)}]{herdeiro1}%
  \BibitemOpen
  \bibfield  {author} {\bibinfo {author} {\bibfnamefont {J.~C.}\ \bibnamefont
  {{Degollado}}}\ and\ \bibinfo {author} {\bibfnamefont {C.~A.~R.}\
  \bibnamefont {{Herdeiro}}},\ }\href {\doibase 10.1007/s10714-013-1598-6}
  {\bibfield  {journal} {\bibinfo  {journal} {General Relativity and
  Gravitation}\ }\textbf {\bibinfo {volume} {45}},\ \bibinfo {pages} {2483}
  (\bibinfo {year} {2013})}\BibitemShut {NoStop}%
\bibitem [{\citenamefont {Sampaio}\ \emph {et~al.}(2014)\citenamefont
  {Sampaio}, \citenamefont {Herdeiro},\ and\ \citenamefont {Wang}}]{herdeiro2}%
  \BibitemOpen
  \bibfield  {author} {\bibinfo {author} {\bibfnamefont {M.~O.~P.}\
  \bibnamefont {Sampaio}}, \bibinfo {author} {\bibfnamefont {C.}~\bibnamefont
  {Herdeiro}}, \ and\ \bibinfo {author} {\bibfnamefont {M.}~\bibnamefont
  {Wang}},\ }\href {\doibase 10.1103/PhysRevD.90.064004} {\bibfield  {journal}
  {\bibinfo  {journal} {Phys. Rev. D}\ }\textbf {\bibinfo {volume} {90}},\
  \bibinfo {pages} {064004} (\bibinfo {year} {2014})}\BibitemShut {NoStop}%
\bibitem [{\citenamefont {Bekenstein}(1973)}]{bekenstein}%
  \BibitemOpen
  \bibfield  {author} {\bibinfo {author} {\bibfnamefont {J.~D.}\ \bibnamefont
  {Bekenstein}},\ }\href@noop {} {\bibfield  {journal} {\bibinfo  {journal}
  {Phys. Rev. D}\ }\textbf {\bibinfo {volume} {7}},\ \bibinfo {pages} {949}
  (\bibinfo {year} {1973})}\BibitemShut {NoStop}%
\bibitem [{\citenamefont {Richartz}\ \emph {et~al.}(2009)\citenamefont
  {Richartz}, \citenamefont {Weinfurtner}, \citenamefont {Penner},\ and\
  \citenamefont {Unruh}}]{mauricio3}%
  \BibitemOpen
  \bibfield  {author} {\bibinfo {author} {\bibfnamefont {M.}~\bibnamefont
  {Richartz}}, \bibinfo {author} {\bibfnamefont {S.}~\bibnamefont
  {Weinfurtner}}, \bibinfo {author} {\bibfnamefont {A.~J.}\ \bibnamefont
  {Penner}}, \ and\ \bibinfo {author} {\bibfnamefont {W.~G.}\ \bibnamefont
  {Unruh}},\ }\href@noop {} {\bibfield  {journal} {\bibinfo  {journal} {Phys.
  Rev. D}\ }\textbf {\bibinfo {volume} {80}},\ \bibinfo {pages} {124016}
  (\bibinfo {year} {2009})}\BibitemShut {NoStop}%
\bibitem [{\citenamefont {Onozawa}\ \emph {et~al.}(1996)\citenamefont
  {Onozawa}, \citenamefont {Mishima}, \citenamefont {Okamura},\ and\
  \citenamefont {Ishihara}}]{ozonawa}%
  \BibitemOpen
  \bibfield  {author} {\bibinfo {author} {\bibfnamefont {H.}~\bibnamefont
  {Onozawa}}, \bibinfo {author} {\bibfnamefont {T.}~\bibnamefont {Mishima}},
  \bibinfo {author} {\bibfnamefont {T.}~\bibnamefont {Okamura}}, \ and\
  \bibinfo {author} {\bibfnamefont {H.}~\bibnamefont {Ishihara}},\ }\href
  {\doibase 10.1103/PhysRevD.53.7033} {\bibfield  {journal} {\bibinfo
  {journal} {Phys. Rev. D}\ }\textbf {\bibinfo {volume} {53}},\ \bibinfo
  {pages} {7033} (\bibinfo {year} {1996})}\BibitemShut {NoStop}%
\end{thebibliography}

%

\end{document}